 \documentclass[journal, 10pt]{IEEEtran}
\usepackage{subfigure}
\usepackage{amsmath}
\usepackage{amsthm}
\usepackage{dsfont}
\usepackage{comment}
\usepackage{graphicx}
\usepackage{amssymb}
\usepackage{epsfig}
\usepackage{authblk}
\usepackage[all]{xy}
\usepackage[bbgreekl]{mathbbol}
\usepackage{pstricks,pst-node,pst-text,pst-3d}
\usepackage{pst-node}
\newpsstyle{Cblue}{fillstyle=solid,fillcolor=blue!30}
\newpsstyle{Cred}{fillstyle=solid,fillcolor=red!30,shadow=true}
\newpsstyle{Cdet}{}
\newpsstyle{Cmisdet}{fillstyle=solid,shadow=true}
\usepackage[square, sort, numbers]{natbib}

\newtheorem{alg}{Algorithm}
\newtheorem{rem}{Remark}
\newtheorem{ex}{Example}

\begin{document}

\title{Estimating the Static Parameters in Linear Gaussian Multiple Target Tracking Models}

\author{Sinan~Y{\i}ld{\i}r{\i}m$^{a, b}$,
        Lan Jiang$^b$,
        Sumeetpal S. Singh$^b$,
        Tom Dean%$^c$
        \thanks{$a$: Statistical Laboratory, Department of Pure Mathematics and Mathematical Statistics, University of Cambridge, UK}
        \thanks{$b$: Department of Engineering, University of Cambridge, UK}
        %\thanks{$c$:  }
        \thanks{L. Jiang, S.S. Singh and T. Dean's research is funded by the Engineering and Physical Sciences Research Council (EP/G037590/1) whose support is gratefully acknowledged.}
        \thanks{Some of the results in this work was presented by the authors in \citet{Yildirim_et_al_2012b} }
        }

%\author[1, 2]{Sinan Y{\i}ld{\i}r{\i}m}
%\author[2]{Lan Jiang}
%\author[2]{Sumeetpal S. Singh}
%\author[2]{Tom Dean}
%
%\affil[1]{Statistical Laboratory, DPMMS, University of Cambridge, UK}
%\affil[2]{Department of Engineering, University of Cambridge, UK}

% \markboth{MANUSCRIPT SUBMITTED TO IEEE TRANSACTIONS ON SIGNAL PROCESSING IN 2012}{}
\markboth{MANUSCRIPT SUBMITTED TO arXiv.org, DECEMBER 2012}{}
% \markboth{Journal of \LaTeX\ Class Files,~Vol.~6, No.~1, January~2007}%
% {Shell \MakeLowercase{\textit{et al.}}: Bare Demo of IEEEtran.cls for Journals}

\maketitle

% \abstract

\begin{abstract}
We present both offline and online maximum likelihood estimation (MLE) techniques for inferring the static parameters of a multiple target tracking (MTT) model with linear Gaussian dynamics. We present the batch and online versions of the expectation-maximisation (EM) algorithm for short and long data sets respectively, and we show how Monte Carlo approximations of these methods can be implemented. Performance is assessed in numerical examples using simulated data for various scenarios and a comparison with a Bayesian estimation procedure is also provided.
\end{abstract}

\section{Introduction}
The multiple target tracking (MTT) problem concerns the analysis of data from multiple moving objects which are
partially observed in noise to extract accurate motion trajectories. The MTT framework has been traditionally applied to solve surveillance problems but more recently there has been a surge of interest in Biological Signal Processing, e.g. see \cite{Yoon_and_Singh_2008}.

The MTT framework is comprised of the following ingredients. A set of multiple independent targets moving in the surveillance region in a Markov fashion. The number of targets varies over time due to departure of existing targets (known as death) and the arrival of new targets (known as birth). The initial number of targets are unknown and the maximum number of targets present at any given time is unrestricted. At each time each target may generate an observation which is a noisy record of its \emph{state}. Targets that do not generate observations are said to be undetected at that time. Additionally, there may be spurious observations generated which are unrelated to targets (known as clutter). The observation set at each time is the collection of all target generated and false measurements recorded at that time, but without any information on the origin or association of the measurements.  False measurements, unknown origin of recorded measurements, undetected targets and a time varying number of targets render the task of extracting the motion trajectory of the underlying targets from the observation record, which is known as \emph{tracking} in the literature, a highly challenging problem.

There is a large body of work on the development of algorithms for tracking multiple moving targets. These algorithms can be categorised by how they handle the data association (or unknown origin of recorded measurements) problem. Among the main approaches are the Multiple Hypothesis Tracking (MHT) algorithm \citep{Reid_1979} and the probabilistic MHT (PMHT) variant \citep{Streit_and_Luginbuhi_1995}, the joint probabilistic data association filter (JPDAF) \citep{Bar-Shalom_and_Fortmann_1988, Bar-Shalom_and_Li_1995},  and the probability hypothesis density (PHD) filter \citep{Mahler_2003, Singh_et_al_2009}. With the advancement of Monte Carlo methodology, sequential Monte Carlo (SMC) (or particle filtering) and Markov chain Monte Carlo (MCMC) methods have been applied to the MTT problem, e.g. SMC and MCMC implementations of JPDA \citep{Ng_et_al_2005, Hue_et_al_2002}, SMC implementations of the MHT and PMHT \citep{Vermaak_et_al_2005, Oh_et_al_2009}, and PHD filter \citep{Vo_et_al_2003, Vo_et_al_2005, Whiteley_et_al_2010}.

Compared to the huge amount of work on developing tracking algorithms, the problem of estimating the static parameters of the tracking model has been largely neglected, although it is rarely the case that these parameters are known. Some exceptions include the work of \citet{Storlie_et_al_2009} where they extended the MHT algorithm to simultaneously estimate the parameters of the MTT model. A full Bayesian approach for estimating the model parameters using MCMC was presented in \citet{Yoon_and_Singh_2008}. \citet{Singh_et_al_2011} presented an approximated maximum likelihood method derived by using a Poisson approximation for the posterior distribution of the hidden targets which is also central to the derivation of PHD filter in \citet{Mahler_2003}. Additionally, versions of PHD and Cardinalised PHD (CPHD) filters that can learn the clutter rate and detection profile  while filtering are proposed in \cite{Mahler_et_al_2011}.

In this paper, we present maximum likelihood estimation (MLE) algorithms to infer \emph{all} the static parameters of the MTT model when the individual
targets move according to a linear Gaussian state-space model and when  the target generated observations are linear functions of the target state corrupted with additive Gaussian noise; we will henceforth call this a linear Gaussian MTT model. We maximise the likelihood function using the expectation-maximisation (EM) algorithm and we present both online and batch EM algorithms. For a linear Gaussian MTT model we are able to present the exact recursions for updating static parameter estimate. To the best of our knowledge, this is a novel development in the target tracking field. We stress though that these recursions are not obvious by virtue of the model being linear Gaussian. This is because the MTT model allows for false measurements, unknown origin of recorded measurements, undetected targets and a time varying number of targets with unknown birth and death times. To implement the proposed EM algorithms, an estimate of the posterior distribution of the hidden targets given the observations is required, and in the linear gaussian setting, the continuous values of the target states can be marginalised out.  But, because the number of possible association of observations to targets grows very quickly with time, we have to resort to approximation schemes that focus the computation in the expectation(E)-step of the EM algorithms on the most likely associations; that is, we approximate the E-step with a Monte Carlo method. For this we employ both SMC and MCMC which give rise to the following different MLE algorithms:
\begin{itemize}
\item SMC-EM and MCMC-EM algorithms for offline estimation; and
\item SMC online EM for online estimation.
\end{itemize}
We implement these three algorithms for simulated examples under various tracking scenarios and provide recommendations for the practitioner on which one is to be preferred.

The EM algorithms we present in this paper can be implemented with any Monte-Carlo scheme for inferring the target states in MTT and reducing the errors in the approximation of the E-step can only be beneficial to the EM parameter estimates. We do not fully explore the use of the various Monte Carlo target tracking algorithms that have been proposed in the literature and instead focus on the following two. When using SMC to approximate the E-step, we compute the $L$-best assignments  \citep{Murty_1968} as the sequential proposal scheme of the particle filter. This $L$-best assignments approached has appeared previously in the literature in the context of tracking, e.g. see \citet{Cox_and_Miller_1995, Ng_et_al_2005, Danchick_and_Newnam_2006}. The MCMC algorithm we use for the E-step is the MCMC-DA algorithm proposed for target tracking in \citet{Oh_et_al_2009}.
For further assessment/comparison of the EM algorithms, we also implement a full Bayesian estimation approach which is essentially a Gibbs like sampler for estimating the static parameters that alternates between sampling the target states and static parameter. Note that the Bayesian approach is not novel and as it been proposed by \citet{Yoon_and_Singh_2008}. It is implemented in this work for the purpose of comparison with the MLE techniques.

The remainder of the paper is organised as follows. In Section \ref{sec: Multiple target tracking model}, we describe the MTT model and formulate the static parameter estimation problem. In Section \ref{sec: EM algorithms for MTT}, we present the batch and online EM algorithms. Section \ref{sec: Experiments and results} contains the numerical examples and we conclude the paper with a discussion of our findings in Section \ref{sec: Conclusion}. The Appendix contains further details on the derivation of the MTT EM algorithm, and details of the SMC and MCMC algorithms we use in this paper.

\subsection{Notation}
We introduce random variables (also sets and mappings) with capital letters such as $X, Y, Z, \mathbf{X}, A$ and denote their realisations by corresponding small case letters $x, y, z, \mathbf{x}, a$. If a non-discrete random variable $X$ has a density $\nu(x)$, with all densities being defined w.r.t.\ the Lebesgue measure (denoted by $dx$), we write $X \sim \nu(\cdot)$ to make explicit the law of $X$. We use $\mathbb{E}_{\theta}[ \cdot | \cdot ]$ for the (conditional) expectation operator; for jointly distributed random variables $X, Y$ and $Z$ and a function $(x, z) \rightarrow f(x, z)$, $\mathbb{E}_{\theta}[f(X, Z) | Y=y]$ is the expectation of the random variable $f(X, Z)$ w.r.t.\ the joint distribution of $X, Z$ conditioned on $Y=y$. $\mathbb{E}_{\theta} [ f(X, z) | y]$ is the expectation of the function $x \rightarrow f(x, z)$ for a fixed $z$ given $Y = y$.

\section{Multiple target tracking model} \label{sec: Multiple target tracking model}
Consider first a \emph{single} target tracking model where a moving object (or target) is observed when it traverses in a surveillance region. We define the target state and the noisy observation at time $t$ to be the random variables $X_{t} \in \mathcal{X} \subset \mathbb{R}^{d_{x}}$ and $Y_{t} \in \mathcal{Y} \subset \mathbb{R}^{d_{y}}$ respectively. The statistical model most commonly used for the evolution of a target and its observations $\{ X_{t}, Y_{t} \}_{t \geq 1}$ is the hidden Markov model (HMM). In a HMM, it is assumed that $\left\{ X_{t} \right\}_{t \geq 1}$  is a hidden Markov process with initial and transition probability densities $\mu_{\psi}$ and $f_{\psi}$, respectively, and $\left\{ Y_{t} \right\}_{t \geq 1}$ is the observation process with the conditional observation density $g_{\psi}$, i.e.
\begin{align} \label{eq: state-space equations}
\begin{split}
&X_{1} \sim \mu_{\psi}(\cdot), \quad X_{t} | (X_{1:t-1} = x_{1:t-1}) \sim f_{\psi} ( \cdot | x_{t-1}) \\
&Y_{t} |  \left( \left\{ X_{i} =  x_{i} \right\}_{i \geq 1}, \left\{ Y_{i} = y_{i} \right\}_{i \neq t} \right) \sim g_{\psi} ( \cdot | x_{t}).
\end{split}
\end{align}
Here the densities $\mu_{\psi}$, $f_{\psi}$ and $g_{\psi}$ are parametrised by a real valued vector $\psi \in \Psi \subset \mathbb{R}^{d_{\psi}}$. In this paper, we consider a specific type of HMM, the Gaussian linear state-space model (GLSSM), which can be specified as
\begin{equation}
\begin{split}
& \mu_{\psi}(x) = \mathcal{N}(x; \mu_{b}, \Sigma_{b}), \quad f_{\psi}(x' | x) = \mathcal{N} (x'; Fx, W), \\
& g_{\psi}(y | x) = \mathcal{N}(y; G x, V). \label{eq: Kalman filter equations}
\end{split}
\end{equation}
where $\mathcal{N}(x; \mu, \Sigma)$ denotes the probability density function for the multivariate normal distribution with mean $\mu$ and covariance $\Sigma$. In this case, $\psi=(\mu_{b}, \Sigma_{b}, F, G, W, V)$.

In a MTT model, the state and the observation at each time ($t \geq 1$) are random finite sets, $\mathbf{X}_{t} = \left( X_{t, 1}, X_{t, 2}, \ldots, X_{t, K^{x}_{t}} \right)$ and $\mathbf{Y}_{t} = \left( Y_{t, 1}, Y_{t, 2}, \ldots, Y_{t, K^{y}_{t}} \right)$. Here each element of $\mathbf{X}_{t}$ is the state of an individual target and elements of $\mathbf{Y}_{t}$ are the distinct measurements of these targets at time $t$. The number of targets $K^{x}_{t}$ under surveillance changes over time due to targets entering and leaving the surveillance region $\mathcal{X}$. $\mathbf{X}_{t}$ evolves to $\mathbf{X}_{t+1}$ as follows: with probability $p_{s}$ each target $\mathbf{X}_{t}$ `survives' and is displaced according to the state transition density $f_{\psi}$ in \eqref{eq: Kalman filter equations}, otherwise it dies. The random deletion and Markov motion happens independently for all the elements of $\mathbf{X}_{t}$. In addition to the surviving targets, new targets are created. The number of new targets created per time follows a Poisson distribution with mean $\lambda_{b}$ and each of their states is initiated independently according to the initial density $\mu_{\psi}$ in \eqref{eq: Kalman filter equations}. Now $\mathbf{X}_{t+1}$ is defined to be the superposition of the states of the surviving and evolved targets from time $t$ and the newly born targets at time $t + 1$. The elements of $\mathbf{X}_{t}$ are observed through a process of random thinning and displacement: with probability $p_{d}$, each point of $\mathbf{X}_{t}$ generates a noisy observation in the observation space $\mathcal{Y}$ through the observation density $g_{\psi}$ in \eqref{eq: Kalman filter equations}. This happens independently for each point of $\mathbf{X}_{t}$. In addition to these target generated observations, false measurements are also generated. The number of false measurements collected at each time follows a Poisson distribution with mean $\lambda_{f}$ and their values are uniform over $\mathcal{Y}$. $\mathbf{Y}_{t}$ is the superposition of observations originating from the detected targets and these false measurements.

\begin{figure*}[t!]
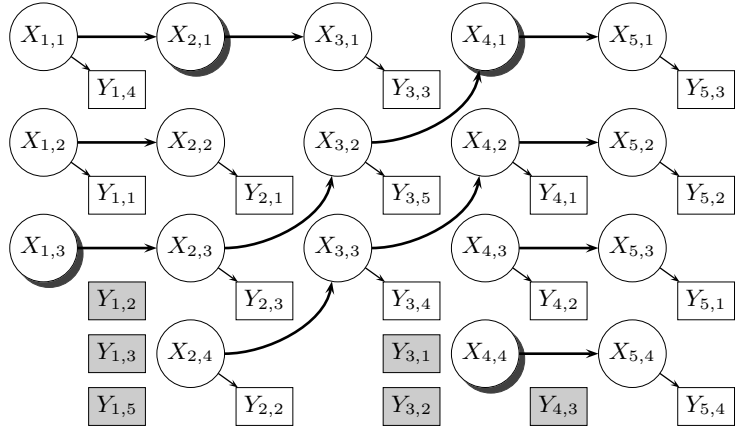

\small
\begin{tabular}{|l|}
\hline
\centerline{\textbf{Complete list of random variables of the MTT model}}  \\
 $X_{t, k}$, $Y_{t, k}$: $k$'th target and $k$'th observation at time $t$.\\
 $\mathbf{X}_{t} = \{ X_{1}, \ldots, X_{K^{x}_{t}} \}$, $\mathbf{Y}_{t} = \{ Y_{t, 1}, \ldots, Y_{t, K^{y}_{t}}\}$: Sets of targets and observations at time $t$. \\
 $K^{b}_{t}, K^{f}_{t}$: Numbers of newborn targets and false measurements at time $t$ \\
 $K^{s}_{t}, K^{d}_{t}$: Numbers of targets survived from time $t-1$ to time $t$ and detected at time $t$. \\
 $K^{x}_{t}, K^{y}_{t}$: Numbers of alive targets and observations at time $t$. $K^{x}_{t} = K^{s}_{t} + K^{b}_{t}$, $K^{y}_{t} = K^{d}_{t} + K^{f}_{t}$. \\
 $C^{s}_{t}$: $K^{x}_{t-1} \times 1$ vector of $0$'s and $1$'s indicating surviving targets from time $t-1$ to time $t$. \\
 $C^{d}_{t}$: $K^{x}_{t} \times 1$ vector of $0$'s and $1$'s indicating detected targets at time $t$. \\
 $I^{s}_{t}$: $K^{s}_{t} \times 1$ vector of labels of surviving targets from time $t-1$ to time $t$. \\
 $I^{d}_{t}$: $K^{d}_{t} \times 1$ vector of labels of detected targets at time $t$. \\
 $A_{t}: \{1, \ldots, K^{d}_{t} \} \rightarrow \{1, \ldots, K^{y}_{t} \}$: Association from detected targets to observations at time $t$. \\
 \hline
 \end{tabular}
\begin{center}
\psset{arrows=->,mnode=circle,linewidth=0.0pt, colsep=0.15cm,rowsep=0.00cm}
\begin{psmatrix}
[name=X11,  style=Cdet] $X_{1, 1}$ & & [name=X21,  style=Cmisdet] $X_{2, 1}$ & & [name=X31,  style=Cdet] $X_{3, 1}$  & & [name=X41,  style=Cmisdet] $X_{4, 1}$  & & [name=X51,  style=Cdet] $X_{5, 1}$  \\
&[name = Y11, mnode=r] \psframebox{$Y_{1, 4}$} & & & & [name=Y31, mnode=r] \psframebox{$Y_{3, 3}$} & & & & [name=Y51, mnode=r] \psframebox{$Y_{5, 3}$} \\
[name=X12,  style=Cdet]$X_{1, 2}$ & & [name=X22,  style=Cdet] $X_{2, 2}$ & & [name=X32,  style=Cdet]$X_{3, 2}$  & & [name=X42, style=Cdet ]$X_{4, 2}$ & & [name=X52,  style=Cdet]$X_{5, 2}$ \\
& [name=Y12, mnode=r] \psframebox{$Y_{1, 1}$} & & [name=Y22, mnode=r] \psframebox{$Y_{2, 1}$} & & [name=Y32, mnode=r] \psframebox{$Y_{3, 5}$}  & & [name=Y42, mnode=r] \psframebox{$Y_{4, 1}$}  & & [name=Y52, mnode=r] \psframebox{$Y_{5, 2}$} \\
[name=X13,  style=Cmisdet]$X_{1, 3}$ & & [name=X23,  style=Cdet]$X_{2, 3}$ & & [name=X33,  style=Cdet]$X_{3, 3}$ & & [name=X43,  style=Cdet]$X_{4, 3}$  & & [name=X53,  style=Cdet]$X_{5, 3}$ \\
& [name=Y13, mnode=r] \psframebox[fillcolor=gray!40,fillstyle=solid]{$Y_{1, 2}$} & & [name=Y23, mnode=r] \psframebox{$Y_{2, 3}$} & & [name=Y33, mnode=r] \psframebox{$Y_{3, 4}$}  & & [name=Y43, mnode=r] \psframebox{$Y_{4, 2}$} & & [name=Y53, mnode=r] \psframebox{$Y_{5, 1}$} \\
& [name=Y15, mnode=r] \psframebox[fillcolor=gray!40,fillstyle=solid]{$Y_{1, 3}$} & [name=X24,  style=Cdet]$X_{2, 4}$ & &  & [name=Y34, mnode=r] \psframebox[fillcolor=gray!40,fillstyle=solid]{$Y_{3, 1}$} & [name=X44,  style=Cmisdet]$X_{4, 4}$ & & [name=X54,  style=Cdet]$X_{5, 4}$ \\
& [name=Y14, mnode=r] \psframebox[fillcolor=gray!40,fillstyle=solid]{$Y_{1, 5}$} & & [name=Y24, mnode=r] \psframebox{$Y_{2, 2}$} & & [name=Y35, mnode=r] \psframebox[fillcolor=gray!40,fillstyle=solid]{$Y_{3, 2}$} & & [name=Y44, mnode=r] \psframebox[fillcolor=gray!40,fillstyle=solid]{$Y_{4, 3}$} & & [name=Y54, mnode=r] \psframebox{$Y_{5, 4}$}
\ncline[linewidth = 1pt]{X11}{X21} \ncline[linewidth = 1pt]{X12}{X22} \ncline[linewidth = 1pt]{X13}{X23}
\ncline[linewidth = 1pt]{X21}{X31} \nccurve[angleA=0,angleB=-100, linewidth = 1pt]{X23}{X32} \nccurve[angleA=0,angleB=-100, linewidth = 1pt]{X24}{X33}
\nccurve[angleA=0,angleB=-100, linewidth = 1pt]{X32}{X41} \nccurve[angleA=0,angleB=-100, linewidth = 1pt]{X33}{X42}
\ncline[linewidth = 1pt]{X41}{X51} \ncline[linewidth = 1pt]{X42}{X52} \ncline[linewidth = 1pt]{X43}{X53} \ncline[linewidth = 1pt]{X44}{X54}
% associations
\ncline[linewidth = 0.5pt]{X11}{Y11} \ncline[linewidth = 0.5pt]{X12}{Y12}
\ncline[linewidth = 0.5pt]{X22}{Y22} \ncline[linewidth = 0.5pt]{X23}{Y23} \ncline[linewidth = 0.5pt]{X24}{Y24}
\ncline[linewidth = 0.5pt]{X31}{Y31} \ncline[linewidth = 0.5pt]{X32}{Y32} \ncline[linewidth = 0.5pt]{X33}{Y33}
\ncline[linewidth = 0.5pt]{X42}{Y42} \ncline[linewidth = 0.5pt]{X43}{Y43} % \ncline[linewidth = 0.5pt]{X44}{Y44}
\ncline[linewidth = 0.5pt]{X51}{Y51} \ncline[linewidth = 0.5pt]{X52}{Y52} \ncline[linewidth = 0.5pt]{X53}{Y53} \ncline[linewidth = 0.5pt]{X54}{Y54}
\end{psmatrix}
\end{center}
\caption{Top: Complete list of the discrete random variables of the MTT model. Bottom: A realisation from MTT model: States of a targets are connected with arrows and with its observations when detected. Undetected targets highlighted with shadows, and false measurements are coloured grey. $C^{s}_{1:5} = \left( \left[  \text{ }\right],  \left[ 1, 1, 1 \right],  \left[ 1, 0, 1, 1 \right],  \left[ 0, 1, 1 \right],  \left[ 1, 1, 1, 1 \right] \right)$; $I^{s}_{1:5} = \left( \left[  \text{ }\right],  \left[ 1, 2, 3 \right],  \left[ 1, 3, 4 \right],  \left[ 2, 3 \right],  \left[ 1, 2, 3, 4 \right] \right)$; $C^{d}_{1:5} = \left( \left[ 1, 1, 0 \right],  \left[ 0, 1, 1, 1 \right],  \left[ 1, 1, 1 \right],  \left[ 0, 1, 1, 0 \right],  \left[ 1, 1, 1, 1 \right] \right)$; $I^{d}_{1:5} = \left( \left[ 1, 2 \right],  \left[ 2, 3, 4 \right],  \left[ 1, 2, 3 \right],  \left[ 2, 3 \right],  \left[ 1, 2, 3, 4 \right] \right)$; $K^{s}_{1:5} = \left( 0, 3, 3, 2, 4 \right)$; $K^{b}_{1:5} = \left( 3, 1, 0, 2, 0 \right)$;  $K^{d}_{1:5} = \left( 2, 3, 3, 2, 4 \right)$;  $K^{f}_{1:5} = \left( 3, 0, 2, 1, 0 \right)$, $A_{1:5} = \left( \left[ 4, 1 \right],   \left[ 1, 3, 2 \right],  \left[ 3, 5, 4 \right],  \left[ 1, 2 \right],  \left[ 3, 2, 1, 4 \right] \right)$.}
\label{fig: MTT}
\hrulefill
\vspace*{3pt}
\end{figure*}

A series of random variables, which are essential for the statistical analysis to follow are now defined. Let $C^{s}_{t}$ be a $K^{x}_{t-1} \times 1$ vector of $1$'s and $0$'s where $1$'s indicate survivals and $0$'s indicate deaths of targets from time $t-1$. For $ i = 1, \ldots, K^{x}_{t-1}$,
\begin{equation}
C^{s}_{t}(i) = \begin{cases}
      1 & \text{$i$'th target at time $t-1$ survives to time $t$} \\
      0 & \text{$i$'th target at time $t-1$ does not survive to $t$}
\end{cases}. \nonumber
\end{equation}
The number of surviving targets at time $t$ is $K^{s}_{t} = \sum_{i = 1}^{K^{x}_{t-1}} C^{s}_{t}(i)$. We also define the $K^{s}_{t} \times 1$ vector $I^{s}_{t}$ containing the indices of surviving targets at time $t$,
\begin{equation}
I^{s}_{t}(i) = \min \left\{ k: \sum_{j = 1}^{k} C^{s}_{t}(j) = i \right\}, \quad i = 1, \ldots, K^{s}_{t}. \nonumber
\end{equation}
Note that $I_{t}^{s}(i)$ will also denote the ancestor of target $i$ from time $t-1$, i.e. $X_{t-1, I^{s}_{t}(i)}$ evolves to $X_{t, i}$ for $i = 1, \ldots, K^{s}_{t}$. Denoting the number of `births' at time $n$ as $K^{b}_{t}$, we have $K^{x}_{t} = K^{s}_{t} + K^{b}_{t}$. Note that according to these definitions, the surviving targets from time $t-1$ are re-labeled as $X_{t, 1}, \ldots, X_{t, K^{s}_{t}}$, and the newly born targets are denoted as $X_{t, K^{s}_{t}+1}, \ldots, X_{t, K^{x}_{t}}$. Next, given $K^{x}_{t}$ targets we define $C^{d}_{t}$ to be a $K^{x}_{t} \times 1$ vector of $1$'s and $0$'s where $1$'s indicate detections and $0$'s indicate non-detections. For $i = 1, \ldots, K^{x}_{t}$,
\begin{equation}
C^{d}_{t}(i) = \begin{cases}
      1 & \text{$i$'th target at time $t$ is detected at time $t$}, \\
      0 & \text{$i$'th target at time $t$ is not detected at time $t$}.
\end{cases}, \nonumber
\end{equation}
Therefore, the number of detected targets at time $t$ is $K^{d}_{t} = \sum_{i = 1}^{K^{x}_{t}} C^{d}_{t}(i)$. Similarly, we also define the $K^{d}_{t}  \times 1$ vector $I^{d}_{t}$ showing the indices of the detected targets,
\begin{equation}
I^{d}_{t}(i) = \min \left\{ k: \sum_{j = 1}^{k} C^{d}_{t}(j) = i \right\}, \quad i = 1, \ldots, K^{d}_{t}. \nonumber
\end{equation}
$I^{d}_{t}(i)$ denotes the label of the $i$-th detected target at time $t$.
So the detected targets at time $t$ are $X_{t, I^{d}_{t}(1)}, \ldots, X_{t, I^{d}_{t}(K^{d}_{t})}$. Finally, defining the number of false measurements at time $t$ as $K^{f}_{t}$, we have $K^{y}_{t} = K^{d}_{t} + K^{f}_{t}$ and the association from the detected targets to the observations can be represented by a one-to-one mapping
\[
A_{t}: \{ 1, \ldots, K^{d}_{t} \} \rightarrow \{ 1, \ldots, K^{y}_{t} \}
\]
where at time $t$ the $i$'th detected target is target $I^{d}_{t}(i)$ with state value $X_{t, I^{d}_{t}(i)}$ and generates $Y_{t, A_{t}(i)}$. We assume that $A_{t}$ is uniform over the set of all $K^{y}_{t}! / K^{f}_{t}!$ possible one-to-one mappings. To summarise, we give the list of the random variables in the MTT model introduced in this section as well as a sample realisation of them in Figure \ref{fig: MTT}.

\normalsize
The main difficulty in an MTT problem is that in general we do not know birth-death times of targets, whether they are detected or not, and which observation point in $\mathbf{Y}_{t}$ is associated to which detected target in $\mathbf{X}_{t}$. Let
\[
Z_{t} = \left( C^{s}_{t}, C^{d}_{t}, K^{b}_{t}, K^{f}_{t}, A_{t} \right)
\]
be the collection of the just mentioned unknown random variables at time $t$, and
\[
\theta = (\psi, p_{s}, p_{d}, \lambda_{b}, \lambda_{f}) \in \Theta = \Psi \times [0, 1]^{2} \times [0, \infty)^{2}
\]
be the vector of the MTT model parameters. We can write the joint likelihood of all the random variables of the MTT model up to time $n$ given $\theta$ as
\[
p_{\theta}(z_{1:n}, \mathbf{x}_{1:n}, \mathbf{y}_{1:n}) =  p_{\theta}(z_{1:n}) p_{\theta}(\mathbf{x}_{1:n} | z_{1:n}) p_{\theta}(\mathbf{y}_{1:n} | \mathbf{x}_{1:n}, z_{1:n})
\]
where
\small
\begin{align}
\begin{split}
& p_{\theta}(z_{1:n}) = \prod_{t = 1}^{n} \Bigg( p_{s}^{k^{s}_{t}} (1 - p_{s})^{k^{x}_{t-1} - k^{s}_{t}}  \mathcal{PO}(k^{b}_{t} ; \lambda_{b})   \\
& \quad \quad \quad \quad \quad \quad\quad\quad\quad\quad \left. p_{d}^{k^{d}_{t}} (1 - p_{d})^{ k^{x}_{t} - k^{d}_{t} }  \mathcal{PO}(k^{f}_{t}; \lambda_{f}) \frac{k^{f}_{t}!}{k^{y}_{t}! } \right) \label{eq: density of z}
\end{split}\\
& p_{\theta}(\mathbf{x}_{1:n} | z_{1:n} ) = \prod_{t = 1}^{n} \left( \prod_{j = 1}^{k^{s}_{t}} f_{\psi}(x_{t, j} | x_{t-1, i^{s}_{t}(j)})  \hspace{-0.2cm} \prod_{j = k^{s}_{t} + 1 }^{k^{x}_{t}}  \hspace{-0.2cm} \mu_{\psi}(x_{t, j}) \right)\label{eq: density of x given z} \\
& p_{\theta}(\mathbf{y}_{1:n} | \mathbf{x}_{1:n}, z_{1:n} ) = \prod_{t = 1}^{n}  \left( \left\vert \mathcal{Y} \right\vert^{-k^{f}_{t}} \prod_{j = 1}^{k^{d}_{t}}  g_{\psi}(y_{t, a_{t}(j)} | x_{t, i^{d}_{t}(j)})  \right) \label{eq: density of y given x and z}
\end{align}
\normalsize
Here $\mathcal{PO}(k; \lambda)$ denotes the probability mass function of the Poisson distribution with mean $\lambda$, $\left\vert \mathcal{Y} \right\vert$ is the volume (w.r.t. the Lebesgue measure) of $\mathcal{Y}$ and the term $k_t^f!/k_t^y!$ in (\ref{eq: density of z}) corresponds to the law of
$A_t.$ The marginal likelihood of the observation sequence $\mathbf{y}_{1:n}$ is
\begin{equation} \label{eq: density of y}
p_{\theta}(\mathbf{y}_{1:n}) = \mathbb{E}_{\theta} \left[ p_{\theta}(\mathbf{y}_{1:n} | \mathbf{X}_{1:n}, Z_{1:n} ) \right].
\end{equation}
The main aim of this paper is, given $\mathbf{Y}_{1:n} = \mathbf{y}_{1:n}$, to estimate the static parameter $\theta^{\ast}$ where we assume the data is generated by some true but unknown $\theta^{\ast} \in \Theta$. Our main contribution is to present the EM algorithms, both batch and online versions, for computing the MLE of $\theta^{\ast}$:
\[
\theta_{\text{ML}} = \arg \max_{\theta \in \Theta} p_{\theta}( \mathbf{y}_{1:n}).
\]
For comparison sake we also present the Bayesian estimate of $\theta^{\ast}$. In the Bayesian approach, the static parameter is treated as random variable taking values $\theta$ in $\Theta$ with a probability density $\eta(\theta)$ and the aim is to evaluate the density of the posterior distribution of $\theta$ given $\mathbf{y}_{1:n}$, i.e.\
\[
p(\theta | \mathbf{y}_{1:n}) = \frac{\eta(\theta) p_{\theta}(\mathbf{y}_{1:n})}{\int_{\Theta} \eta(\theta) p_{\theta}(\mathbf{y}_{1:n}) d \theta}.
\]
\citet{Yoon_and_Singh_2008} use MCMC to sample from $p(\theta | \mathbf{y}_{1:n})$ which integrates both Metropolis-Hastings and Gibbs moves.

\section{EM algorithms for MTT} \label{sec: EM algorithms for MTT}
In this section we present the batch and online EM algorithms for linear Gaussian MTT models. The notation is involved and we provide a list of the important variables used in the derivation of the EM algorithms in Table  \ref{table: EM variables} at the end of the section.

\subsection{Batch EM for MTT} \label{sec: Batch EM for MTT}

Given $\mathbf{Y}_{1:n} = \mathbf{y}_{1:n}$, the EM algorithm for maximising $p_{\theta}(\mathbf{y}_{1:n})$ in \eqref{eq: density of y} is given by the following iterative procedure: if $\theta_{j}$ is the estimate of the EM algorithm at the $j$'th iteration, then at iteration $j + 1$ the estimate is updated by first calculating the following intermediate optimisation criterion, which is known as the expectation (E) step,
\begin{align} \label{eq:EM intermediate quantity}
\begin{split}
Q(& \theta_{j}, \theta) = \mathbb{E}_{\theta_{j}} \left[ \log p_{\theta}(\mathbf{X}_{1:n}, Z_{1:n}, \mathbf{y}_{1:n}) | \mathbf{y}_{1:n} \right]  \\
& = \mathbb{E}_{\theta_{j}} \left[ \log p_{\theta}(Z_{1:n}) + \log p_{\theta}(\mathbf{X}_{1:n}, \mathbf{y}_{1:n} | Z_{1:n} ) | \mathbf{y}_{1:n} \right]  \\
&  = \mathbb{E}_{\theta_{j}} \left[ \log p_{\theta}(Z_{1:n}) \right. \\
&  \quad \left. +\mathbb{E}_{\theta_{j}} \left\{ \log p_{\theta}(\mathbf{X}_{1:n}, \mathbf{y}_{1:n} | Z_{1:n} ) | \mathbf{y}_{1:n}, Z_{1:n}  \right\} | \mathbf{y}_{1:n}\right]
\end{split}
\end{align}
The updated estimate is then computed in the maximisation (M) step
\[
\theta_{j+1} = \arg \max_{\theta \in \Theta} Q(\theta_{j}, \theta)
\]
This procedure is repeated until $\theta_{j}$ converges (or in practice ceases to change significantly). From equations \eqref{eq: Kalman filter equations}-\eqref{eq: density of y given x and z}, it can be shown that the E-step at the $j$'th iteration reduces to calculating the expectations of fifteen sufficient statistics of $\mathbf{x}_{1:n}$, $z_{1:n}$ and $\mathbf{y}_{1:n}$ denoted by $S_{1, n}, \ldots, S_{15, n}$. (From now on, any dependancy on $\mathbf{y}_{1:n}$ in these sufficient statistics and further variables arising from them will be omitted from the notation for simplicity.) Sufficient statistics $S_{1,  n}(\mathbf{x}_{1:n}, z_{1:n})$ to  $S_{7,  n}(\mathbf{x}_{1:n}, z_{1:n})$ are:
\begin{align}
& \sum_{t = 1}^{n} \sum_{k = 1}^{k_{t}^{d}} x_{t, i^{d}_{t}(k)} x_{t, i^{d}_{t}(k)}^{T},  \quad \sum_{t = 1}^{n} \sum_{k = 1}^{k_{t}^{d}} x_{t, i^{d}_{t}(k)} y_{t, a_{t}(k)}^{T}, \nonumber \\
& \sum_{t = 2}^{n} \sum_{k = 1}^{k_{t}^{s}}  x_{t-1, i^{s}_{t}(k)} x_{t-1, i^{s}_{t}(k)}^{T},  \quad \sum_{t = 2}^{n} \sum_{k = 1}^{k_{t}^{s}} x_{t, k} x_{t, k}^{T}, \label{eq: KF sufficient statistics for MTT} \\
& \sum_{t = 2}^{n} \sum_{k = 1}^{k_{t}^{s}}  x_{t-1, i^{s}_{t}(k)} x_{t, k}^{T}, \quad \sum_{t = 1}^{n} \sum_{k = k_{t}^{s} + 1}^{k^{x}_{t}} x_{t, k} , \quad \sum_{t = 1}^{n} \sum_{k = k_{t}^{s} + 1}^{k^{x}_{t}}  x_{t, k} x_{t, k}^{T} \nonumber
\end{align}
These sufficient statistics are related to those used for estimating the static parameters of a linear Gaussian single target tracking model, and this relation will be made more explicit later. The rest of the sufficient statistics $S_{8,  n}(z_{1:n})$ to  $S_{15,  n}(z_{1:n})$ do not depend on $\mathbf{x}_{1:n}$.
\begin{align}
& \left[ S_{8, n}, \ldots, S_{15, n} \right](z_{1:n}) \nonumber \\
& \quad = \sum_{t = 1}^{n} \left[ \sum_{k = 1}^{k_{t}^{d}} y_{t, a_{t}(k)} y_{t, a_{t}(k)}^{T}, k_{t}^{d}, k^{x}_{t},  k^{s}_{t}, k^{x}_{t-1}, k^{b}_{t}, k^{f}_{t}, 1 \right] \label{eq: MTT sufficient statistics}
\end{align}
Let $S_{m, n}^{\theta}$ denote the expectation of the $m$'th sufficient statistic w.r.t.\ the law of the latent variables $\mathbf{X}_{1:n}$ and $Z_{1:n}$ conditional upon the observation $\mathbf{y}_{1:n}$ for a given $\theta$, i.e.
\begin{align}
S_{m, n}^{\theta}  = \begin{cases}
      \mathbb{E}_{\theta}\left[ \left. S_{m, n} \left( \mathbf{X}_{1:n}, Z_{1:n} \right) \right\vert \mathbf{y}_{1:n} \right] & 1 \leq m \leq 7, \\
      \mathbb{E}_{\theta}\left[ \left. S_{m, n} \left( Z_{1:n} \right)  \right\vert \mathbf{y}_{1:n} \right] & 8 \leq m \leq 15.
\end{cases}
\end{align}
Then the solution to the M-step is given by a known function $\Lambda: \left\{ \left( S_{1, n}^{\theta}, \ldots, S_{15, n}^{\theta} \right) \right\} \rightarrow \Theta$ such that at iteration $j$
\begin{equation}
\theta_{j+1} = \arg \max_{\theta} Q(\theta_{j}, \theta) = \Lambda \left( S_{1, n}^{\theta_{j}}, \ldots, S_{15, n}^{\theta_{j}} \right). \nonumber
\end{equation}
The explicit expression of $\Lambda$ depends on the parametrisation of the MTT model, in particular on the parametrisation of the matrices $F, G, W, V, \mu_{b}, \Sigma_{b}$ as in the following example.

\begin{ex} \label{ex: The constant velocity model}
(\textit{The constant velocity model:}) Each target has a position and velocity in the $xy$-plane and hence
\[
X_{t} = \left[ X_{t}(1), X_{t}(2), X_{t}(3), X_{t}(4) \right]^{T} \in \mathcal{X} =  \mathbb{R}^{2} \times  [0, \infty)^{2},
\]
where $X_{t}(1), X_{t}(2)$ are the $x$ and $y$ coordinates and $X_{t}(3), X_{t}(4)$ are the velocities in $x$ and $y$ directions.  Only a noisy measurement of the position of the target is available
\[
\left[ Y_{t}(1), Y_{t}(2) \right] \in \mathcal{Y} = [-\kappa, \kappa]^{2}.
\]
We assumed a bounded $\mathcal{Y}$ and regard observations that are not recorded due to being outside this interval as
also a missed detection.
With reference to \eqref{eq: Kalman filter equations}, the single target state-space model is
\begin{gather}
\mu_{b} = \left[ \mu_{bx}, \mu_{by},  0, 0 \right]^{T}, \quad \Sigma_{b} = \left( \begin{array}{cc} \sigma_{bp}^{2} I_{2 \times 2} & \mathbf{0}_{2 \times 2} \\ \mathbf{0}_{2 \times 2} & \sigma_{bv}^{2} I_{2 \times 2} \end{array} \right)\nonumber\\
F = \left(\begin{array}{cc} I_{2 \times 2} & \Delta I_{2 \times 2} \\ \mathbf{0}_{2 \times 2} & I_{2 \times 2} \end{array}\right), \quad G = \left(\begin{array}{cc} I_{2 \times 2} & \mathbf{0}_{2 \times 2} \end{array}\right)\nonumber\\
W = \left(\begin{array}{cc} \sigma_{xp}^{2} I_{2 \times 2}  &  \mathbf{0}_{2 \times 2} \\ \mathbf{0}_{2 \times 2} & \sigma_{xv}^{2} I_{2 \times 2} \end{array}\right), \quad V = \sigma_{y}^{2} I_{2 \times 2}\nonumber
\end{gather}

Therefore, the parameter vector of this MTT model is
\[
\theta = \left( \lambda_{b}, \lambda_{f}, p_{d}, p_{s}, \mu_{bp}, \mu_{bv}, \sigma_{bp}^{2}, \sigma_{bv}^{2}, \sigma_{xp}^{2}, \sigma_{xv}^{2}, \sigma_{y}^{2} \right).
\]
The update rule $\Lambda$ for $\theta$ at the M-step of the EM algorithm is
\begin{align}
& \mu_{bx} = S_{6, n}^{\theta}(1) / S_{13, n}^{\theta}, \quad \mu_{by} = S_{6, n}^{\theta}(2) / S_{13, n}^{\theta}, \nonumber \\
& \sigma_{bp}^{2} =\frac{1}{2}\;S_{13, n}^{\theta} \text{tr} \left( \left( S_{7, n}^{\theta} - 2 S_{6, n}^{\theta} \mu_{b}^{T} + S_{13, n}^{\theta}  \mu_{b}  \mu_{b}^{T} \right) M_{p}^{T} M_{p} \right)  \nonumber \\
& \sigma_{bv}^{2} = \frac{1}{2}\;S_{13, n}^{\theta} \text{tr} \left( \left( S_{7, n}^{\theta} - 2 S_{6, n}^{\theta} \mu_{b}^{T} + S_{13, n}^{\theta}  \mu_{b}  \mu_{b}^{T} \right) M_{v}^{T} M_{v} \right) \nonumber \\
& \sigma_{xp}^{2} = \text{tr} \left( S_{4, n}^{\theta} M_{p}^{T} M_{p} - 2 S_{5, n}^{\theta} M_{p} F_{p} + S_{3, n}^{\theta}  F_{p}^{T}F_{p}\right) / 2 S_{11, n}^{\theta}, \nonumber \\
& \sigma_{xv}^{2} = \text{tr} \left( S_{4, n}^{\theta} M_{v}^{T} M_{v} - 2 S_{5, n}^{\theta} M_{v} F_{v} + S_{3, n}^{\theta}  F_{v}^{T}F_{v}\right) / 2 S_{11, n}^{\theta}, \nonumber \\
& \sigma_{y}^{2} = \text{tr} \left( S_{8, n}^{\theta} - 2 G S_{2, n}^{\theta} + G S_{1, n}^{\theta}  G^{T} \right) / 2 S_{9, n}^{\theta}, \nonumber \\
& p_{d}= S_{9, n}^{\theta} / S_{10, n}^{\theta} , \quad p_{s}  =  S_{11, n}^{\theta} / S_{12, n}^{\theta},\nonumber\\
& \lambda_{b} = S_{13, n}^{\theta} / S_{15, n}^{\theta}, \quad \lambda_{f} = S_{14, n}^{\theta} / S_{15, n}^{\theta}, \nonumber
\end{align}
where $ M_{p} = \begin{bmatrix} I_{2 \times 2} & 0_{2 \times 2} \end{bmatrix}, M_{v} = \begin{bmatrix} 0_{2 \times 2} & I_{2 \times 2} \end{bmatrix}$, and $F_{p}$ and $F_{v}$ are the upper and lower halves of $F$, that is $F_{p}(i, j) = F(i, j)$ and $F_{v}(i, j) = F(2  + i, j)$ for $i = 1, 2 $ and $j = 1, \ldots, 4$.
\end{ex}

\subsubsection{Estimation of sufficient statistics} \label{sec: Estimation of sufficient statistics}
It is easy to calculate the expectation of the sufficient statistics in \eqref{eq: MTT sufficient statistics} that do not depend on $\mathbf{x}_{1:n}$. Noting that $Z_{t}$ is discrete, we simply calculate $S_{m, n}(z_{1:n})$ for every $z_{1:n}$ with a positive mass w.r.t.\ to the density $p_{\theta}(z_{1:n} | \mathbf{y}_{1:n})$ and calculate the expectations as
\[
S_{m, n}^{\theta} = \sum_{z_{1:n}} S_{m, n}(z_{1:n}) p_{\theta}(z_{1:n} | \mathbf{y}_{1:n}).
\]
For those sufficient statistics in \eqref{eq: KF sufficient statistics for MTT} that depend on $\mathbf{x}_{1:n}$, consider the
last expression in \eqref{eq:EM intermediate quantity} with the following factorisation of the posterior
\begin{equation*} \label{eq: posterior of x and z given y factorized}
p_{\theta}(\mathbf{x}_{1:n}, z_{1:n} | \mathbf{y}_{1:n}) = p_{\theta}(\mathbf{x}_{1:n} | z_{1:n}, \mathbf{y}_{1:n}) p_{\theta}(z_{1:n} | \mathbf{y}_{1:n}).
\end{equation*}
This factorisation suggests that we can write the required expectations as
\begin{align}
S_{m, n}^{\theta} &= \mathbb{E}_{\theta} \left[ \left. S_{m, n}(\mathbf{X}_{1:n}, Z_{1:n} ) \right\vert \mathbf{y}_{1:n} \right] \nonumber \\
& = \mathbb{E}_{\theta} \left[ \left. \mathbb{E}_{\theta} \left[ \left.  S_{m, n}(\mathbf{X}_{1:n}, Z_{1:n}) \right\vert Z_{1:n}, \mathbf{y}_{1:n} \right] \right\vert \mathbf{y}_{1:n} \right]. \label{eq: nested expectation for x dependant sufficient statistics}
\end{align}
Let us define the integrand of the outer expectation in \eqref{eq: nested expectation for x dependant sufficient statistics} which is the conditional expectation
\begin{equation*} \label{eq: inner expectation}
\widetilde{S}_{m, n}^{\theta}(z_{1:n}) = \mathbb{E}_{\theta} \left[ \left.  S_{m, n}(\mathbf{X}_{1:n}, z_{1:n}) \right\vert z_{1:n}, \mathbf{y}_{1:n} \right].
\end{equation*}
as a matrix-valued function with domain $\mathcal{Z}^{n}$. Then, we can obtain $S_{m, n}^{\theta}$ by calculating $\widetilde{S}_{m, n}^{\theta}(z_{1:n})$ for every $z_{1:n}$ with a positive mass w.r.t.\ the density $p_{\theta}(z_{1:n} | \mathbf{y}_{1:n})$ and then calculate
\[
S_{m, n}^{\theta} = \sum_{z_{1:n}} \widetilde{S}_{m, n}^{\theta}(z_{1:n}) p_{\theta}(z_{1:n} | \mathbf{y}_{1:n}).
\]
The crucial point here is that it is possible to calculate $\widetilde{S}_{m, n}^{\theta}(z_{1:n})$ for any given $z_{1:n}$. In fact, the availability of this calculation is based on the following fact: \textit{conditional on $\left\{ Z_{t} \right\}_{t \geq 1}$, $\left\{ \mathbf{X}_{t}, \mathbf{Y}_{t} \right\}_{t \geq 1}$ may be regarded as a collection of independent GLSSMs (with different starting and ending times, possible missing observations) and observations which are not relevant to any of these GLSSMs}. In the context of MTT, each GLSSM corresponds to a target and irrelevant observations correspond to false measurements. We defer details on how $\widetilde{S}_{m, n}^{\theta}(z_{1:n})$ is calculated to Section \ref{sec: Online EM for MTT}.

\subsubsection{Stochastic versions of EM} \label{sec: Stochastic versions of EM}
For exact calculation of the E-step of the EM algorithm we need $p_{\theta}(z_{1:n} | \mathbf{y}_{1:n})$
which is infeasible to calculate due to the huge cardinality of $\mathcal{Z}^{n}$. We thus resort to Monte Carlo approximations of $p_{\theta}(z_{1:n} | \mathbf{y}_{1:n})$ which we then use in the E-step; in literature this approach is generically known as the stochastic EM algorithm \citep{Celeux_and_Diebolt_1985, Wei_and_Tanner_1990, Delyon_et_al_1999}). We know from the previous sections that given $Z_{1:n} = z_{1:n}$ the posterior distribution $p_{\theta}(\mathbf{x}_{1:n} | \mathbf{y}_{1:n}, z_{1:n})$ is Gaussian and conditional expectations can be evaluated. Therefore, it is sufficient to have the Monte Carlo particle approximation for $p_{\theta}(z_{1:n} | \mathbf{y}_{1:n})$ only, which is expressed as
\begin{equation} \label{eq: MC approximation to z given y}
\widehat{p}_{\theta}(z_{1:n} | \mathbf{y}_{1:n}) = \sum_{i = 1}^{N} w_{n}^{(i)} \delta_{z_{1:n}^{(i)}} (z_{1:n}), \quad \sum_{i = 1}^{N} w_{n}^{(i)} = 1.
\end{equation}
Then, the corresponding particle approximations for the expectations of the sufficient statistics are
\begin{align*}
\widehat{S}_{m, n}^{\theta} = \begin{cases}
      \sum_{i = 1}^{N} w_{n}^{(i)}  \widetilde{S}_{m, n}^{\theta} (  z_{1:n}^{(i)} ), & 1 \leq m \leq 7, \\
       \sum_{i = 1}^{N} w_{n}^{(i)} S_{m, n}( z_{1:n}^{(i)} ), & 8 \leq m \leq 15.
\end{cases}
\end{align*}
When $\theta$ changes with each EM iteration, the appropriate update scheme at iteration $j$ involves a stochastic approximation procedure where in the E-step one calculates a weighted average of $\widehat{S}_{m, n}^{\theta_{1}}, \ldots, \widehat{S}_{m, n}^{\theta_{j}}$; the resulting algorithm is known as the stochastic approximation EM (SAEM) \citep{Delyon_et_al_1999}. Specifically, let $\gamma = \left\{ \gamma_{j} \right\}_{j \geq 1}$, called the step-size sequence, be a positive decreasing sequence satisfying
\[
\sum_{j} \gamma_{j}  = \infty, \quad \sum_{j} \gamma_{j}^{2} < \infty.
\]
A common choice is $\gamma_{j} = j^{-\alpha}$ for $0.5 < \alpha \leq 1$. The SAEM algorithm is given in Algorithm \ref{alg: SAEM for the MTT model}.
\small
\begin{alg} \label{alg: SAEM for the MTT model}
\textbf{The SAEM algorithm for the MTT model} \\
Start with $\theta_{1}$ and $\widehat{S}_{\gamma, m, n}^{(0)} = 0$ for $m = 1, \ldots, 15$. For $j = 1, 2, \ldots$
\begin{itemize}
\item \textbf{E-step:} Calculate $\widehat{S}_{m, n}^{\theta_{j}}$ for each $m$, and then calculate the weighted averages
\begin{equation} \label{eq: E-step of SAEM}
\widehat{S}_{\gamma, m, n}^{(j)} = \left( 1 - \gamma_{j} \right)  \widehat{S}_{\gamma, m, n}^{(j-1)} + \gamma_{j} \widehat{S}_{m, n}^{\theta_{j}}.
\end{equation}
\item \textbf{M-step} Update the parameter estimate using $\Lambda(\cdot)$ as before
\[
\theta_{j+1} = \Lambda \left( \widehat{S}_{\gamma, 1, n}^{(j)}, \ldots, \widehat{S}_{\gamma, 15, n}^{(j)} \right).
\]
\end{itemize}
\end{alg}
\normalsize
In general, the Monte Carlo approximation $\widehat{p}_{\theta_{j}}(z_{1:n} | \mathbf{y}_{1:n})$ in \eqref{eq: E-step of SAEM} is performed either sampling $N$ samples from $p_{\theta_{j}}(z_{1:n} | \mathbf{y}_{1:n})$ using a MCMC method (in which case weights $w_{n}^{(i)} = 1/N$, $i = 1, \ldots, N$) or using a SMC method with $N$ particles. Depending on which method is used, we will call the resulting algorithm MCMC-EM or SMC-EM, respectively. For MCMC, we use the MCMC-DA algorithm of  \cite{Oh_et_al_2009}, but with some refinements of the MCMC proposals. (Details are available from the authors.)

We use SMC to obtain the approximations $\left\{\widehat{p}_{\theta}(z_{1:t} | \mathbf{y}_{1:t}) \right\}_{1 \leq t \leq n}$ sequentially as follows. Assume that we have the approximation at time $t-1$
\[
\widehat{p}_{\theta}(z_{1:t-1} | \mathbf{y}_{1:t-1}) = \sum_{i = 1}^{N} w_{t-1}^{(i)} \delta_{z_{1:t-1}^{(i)}} (z_{1:t-1}).
\]
To avoid weight degeneracy, at each time one can resample from $\widehat{p}_{\theta}(z_{1:t-1} | \mathbf{y}_{1:t-1}) $ to obtain a new collection of $N$ particles and then proceed to the time $t$. Alternatively, this resampling operation can be done according to a criterion which measures the weight degeneracy (e.g. see \citet{Doucet_et_al_2000}). We define the $N \times 1$ random mapping
\[
\Pi_{t}: \{ 1, \ldots, N\} \rightarrow \{ 1, \ldots, N \}
\]
containing the indices of the resampled particles, i.e. $\Pi_{t}(i) = j$ if the $i$'th resampled particle is $z_{1:t-1}^{(j)}$. (If no resampling is performed at the end of time $t-1$, then $\Pi_{t}(i) = i$ for all $i$.) Then, given $\mathbf{y}_{t}$ and $\Pi_{t} = \pi_{t}$, the particle $z_{t}^{(i)}$ at time $t$ is sampled from a proposal distribution
\[
q_{\theta} \left( z_{t} \left\vert z_{1:t-1}^{(\pi_{t}(i))}, \mathbf{y}_{1:t} \right. \right)
\]
 for $i = 1, \ldots, N$. Therefore, $z_{t}^{(i)}$ is connected to $z_{1:t-1}^{(\pi_{t}(i))}$ and the $i$'th path particle at time $t$ is $z_{1:t}^{(i)} = ( z_{t}^{(i)}, z_{1:t-1}^{(\pi_{t}(i))} )$ and its new weight is
\begin{equation} \label{eq: particle weight}
w_{t}^{(i)} \propto \bar{w}_{t-1}^{(\pi_{t}(i))} \times \frac{p_{\theta} ( z_{t}^{(i)} \vert z_{t-1}^{(\pi_{t}(i))} ) p_{\theta} ( \mathbf{y}_{t} \vert \mathbf{y}_{1:t-1}, z_{1:t}^{(i)} )}{q_{\theta} ( z_{t}^{(i)} \vert z_{1:t-1}^{(\pi_{t}(i))}, \mathbf{y}_{1:t} )}
\end{equation}
where, for $i = 1, \ldots, N$, we take $\bar{w}_{t-1}^{(i)} = 1/N$ if resampling is performed and $\bar{w}_{t-1}^{(i)} = w_{t-1}^{(i)}$ otherwise.

Note that we also need to implement SMC for the online EM algorithm in order to obtain a Monte Carlo approximation of the E-step. Our SMC algorithm calculates the $L$-best linear assignments \citep{Murty_1968} as the sequential proposal; see Appendix \ref{sec: SMC algorithm for MTT} for details.

\subsection{Online EM for MTT} \label{sec: Online EM for MTT}

We showed in the previous section how to implement the batch EM algorithm for MTT using Monte Carlo approximations. However, the batch EM algorithm is computationally demanding when the data sequence $\mathbf{y}_{1:n}$ is long since one iteration of the EM requires a complete browse of the data. In these situations, the online version of the EM algorithm which updates the parameter estimates as a new data record is received at each time can be a much cheaper alternative. In this section, we present a SMC online EM algorithm for linear Gaussian MTT models.

An important observation at this point is that the sufficient statistics of interest for the EM algorithm have a certain additive form such that the difference of $S_{m, n}(\mathbf{x}_{1:n}, z_{1:n})$ and $S_{m, n-1}(\mathbf{x}_{1:n-1}, z_{1:n-1})$ only depends on $(\mathbf{x}_{n-1}, \mathbf{x}_{n}, \mathbf{y}_{n} )$. This enables us to compute the required expectations in the E-step of the EM algorithm effectively in an online manner. We shall see in this section that, with a fixed amount of computation and memory per time, it is possible to update from $\widetilde{S}_{m, t-1}^{\theta}(z_{1:t-1})$ to $\widetilde{S}_{m, t}^{\theta}(z_{1:t})$ given $\mathbf{y}_{t}$ and $z_{t}$ at time $t$. To show how to handle the sufficient statistics in \eqref{eq: KF sufficient statistics for MTT} for the MTT model, we first start with a single GLSSM and then extend the idea to the MTT case by showing the relation between the sufficient statistics in a single GLSSM and in the MTT model.

\subsubsection{Online smoothing in a single GLSSM} \label{sec: Online smoothing in a single GLSSM}

Consider the HMM $\left\{ X_{t}, Y_{t} \right\}_{t \geq 1}$ defined in \eqref{eq: state-space equations}. It is possible to evaluate expectations of additive functionals of $X_{1:n}$ of the form
\[
S_{n}(x_{1:n}) = s(x_{1}) + \sum_{t = 2}^{n} s(x_{t-1}, x_{t})
\]
(with possible dependancy on $y_{1:n}$ also allowed) w.r.t.\ the posterior density $p_{\theta}(x_{1:n} | y_{1:n})$ in an online manner using only the filtering densities $\{ p_{\theta}(x_{t} | y_{1:t}) \}_{1 \leq t \leq n}$. The technique is based on the following recursion on the intermediate function \citep{Del_Moral_et_al_2009, Cappe_2011}
\begin{align}
T_{t}^{\theta}(x_{t}) :=& \mathbb{E}_{\theta} \left[ S_{t}(X_{1:t}) | X_{t} = x_{t}, y_{1:t} \right] \nonumber \\
 =& \mathbb{E}_{\theta} \left[ \left. T_{t-1}^{\theta}(X_{t-1}) + s(X_{t-1}, x_{t}) \right\vert y_{1:t-1}, x_{t} \right]  \label{eq:FSintermediate}
\end{align}
with the initial condition $T_{1}^{\theta}(x_{1}) = s(x_{1})$. Note that the expectation required for the recursion is w.r.t.\ the backward transition density  $p_{\theta}(x_{t-1} | y_{1:t-1}, x_{t})$. The required expectation $\mathbb{E}_{\theta} \left[ S_{n}(X_{1:n}) | y_{1:n} \right]$ can then be calculated as the expectation of the intermediate function $T_{n}^{\theta}(x_{n})$ w.r.t.\ the filtering density $p_{\theta}(x_{n} | y_{1:n})$, that is,
\[
\mathbb{E}_{\theta} \left[ \left. S_{n}(X_{1:n}) \right\vert y_{1:n} \right] = \mathbb{E}_{\theta} \left[ \left. T_{n}^{\theta}(X_{n}) \right\vert y_{1:n} \right].
\]
Consider now the GLSSM that is defined in \eqref{eq: Kalman filter equations}, where, additionally, $Y_{t}$ is possibly missing/undetected and $C_{t}^{d}$ is the indicator of detection at time $t$. It is well known that, given $\{ (Y_{t}, C_{t}^{d})  = (y_{t}, c_{t}^{d}) \}_{t \geq 1}$, the prediction and filtering densities  $p_{\theta}(x_{t} | y_{1:t-1}, c_{1:t-1}^{d})$ and $p_{\theta}(x_{t} | y_{1:t}, c_{1:t}^{d})$ are Gaussians with means  $\left( \mu_{t|t-1}, \mu_{t|t} \right)$ and covariances $\left( \Sigma_{t|t-1}, \Sigma_{t|t} \right)$ and are updated sequentially as follows:
\begin{align}
&  (\mu_{t | t-1}, \Sigma_{t | t-1}) = F \mu_{t-1 | t-1},  F  \Sigma_{t | t-1} F^{T} + W, \label{eq: GLSSM prediction update} \\
&  (\mu_{t | t}, \Sigma_{t | t}) \hspace{-0.1cm}=  \hspace{-0.1cm}
\begin{cases}
	\begin{aligned}
		&  \hspace{-0.2cm} \left( \mu_{t | t-1} + \Sigma_{t | t-1} G^{T} \Gamma_{t}^{-1} \epsilon_{t}, \right.\\
		&  \hspace{-0.2cm} \left.\Sigma_{t | t-1}  - \Sigma_{t | t-1} G^{T} \Gamma_{t}^{-1} G \Sigma_{t | t-1}  \right), \\
	\end{aligned} & \hspace{-0.4cm} c_{t}^{d} = 1 \\
	\left( \mu_{t | t-1}, \Sigma_{t | t-1} \right),  & \hspace{-0.4cm} c_{t}^{d} = 0.
\end{cases} \hspace{-0.1cm}\label{eq: GLSSM filter update}
\end{align}
where $\Gamma_{t} = G \Sigma_{t | t-1} G^{T} + V$ and $\epsilon_{t} = y_{t} - G \mu_{t | t-1}$. Also, letting $B_{t} = \Sigma_{t|t} F^{T} (F \Sigma_{t|t} F^{T} + W)^{-1} $, $b_{t} = (I_{d_{x} \times d_{x}} - B_{t} F ) \mu_{t|t}$, and $\Sigma_{t|t+1} = (I_{d_{x} \times d_{x}} - B_{t} F ) \Sigma_{t|t}$ we can show that the backward transition density required for the forward smoothing recursion (\ref{eq:FSintermediate}) is Gaussian as well
\[
p_{\theta}(x_{t-1}| y_{1:t-1}, c_{1:t-1}^{d}, x_{t}) =\mathcal{N} \left(x_{t-1}; B_{t-1} x_{t} + b_{t-1}, \Sigma_{t-1 | t} \right). \nonumber
\]
We define the matrix valued functions
\[
\bar{S}_{m, l}: \mathcal{X}^{l} \times \{0, 1\}^{l} \times \mathcal{Y}^{l} \rightarrow \mathbb{R}^{d_{x} \times d_{m}},
\]
such that $\bar{S}_{m, l}(x_{1:l}, c^{d}_{1:l}, y_{1:l} )$ for $ m = 1, \ldots, 7$ are in the following form:
\begin{align}
\begin{split}
&  \sum_{t = 1}^{l} c_{t}^{d} x_{t} x_{t}^{T},  \quad \sum_{t = 1}^{l} c_{t}^{d} x_{t} y_{t}^{T},  \quad \sum_{t = 2}^{l} x_{t-1} x_{t-1}^{T}, \\
& \sum_{t = 2}^{l} x_{t} x_{t}^{T}, \quad \sum_{t = 2}^{l} x_{t-1} x_{t}^{T}, \quad x_{1}, \quad x_{1} x_{1}^{T}.
\end{split}  \label{eq: KF sufficient statistics}
\end{align}
(so, $d_{2} = d_{y}$ and $d_{6} = 1$, else $d_{m} = d_{x}$). These functions are actually the sufficient statistics in the MTT model corresponding to a single target. Then it is possible to define the incremental functions
\begin{equation}
\bar{s}_{m}: \left( \mathcal{X} \cup \mathcal{X}^{2} \right) \times \{0, 1\} \times \mathcal{Y} \rightarrow \mathbb{R}^{d_{x} \times d_{m}}
\label{eq:barsm}
\end{equation}
where $\bar{s}_{m}$'s are defined such that for $m = 1, \ldots, 7$
\[
\bar{S}_{m, l}(x_{1:l}, c_{1:l}^{d}, y_{1:l}) = \bar{s}_{m}(x_{1}, c_{1}^{d}, y_{1}) +  \sum_{t = 2}^{l} \bar{s}_{m}(x_{t-1}, x_{t}, c_{t}^{d}, y_{t}).
\]
For example, $\bar{s}_{1}(x_{1}, c_{1}^{d}, y_{1}) = c_{1}^{d} x_{1} x_{1}^{T}$, $\bar{s}_{3}(x_{1}, c_{1}^{d}, y_{1}) = 0_{d_{x} \times d_{x}}$, $\bar{s}_{5}(x_{t-1}, x_{t}, c_{t}^{d}, y_{t}) = x_{t-1} x_{t}^{T}$, $\bar{s}_{6}(x_{1}, c_{1}^{d}, y_{1}) = x_{1}$, $\bar{s}_{7}(x_{t-1}, x_{t}, c_{t}^{d}, y_{t}) = 0_{d_{x} \times d_{x}}$, etc. We observe that each sufficient statistic is a matrix valued quantity, hence its expectation can be calculated using forward smoothing by treating each element of the matrix separately. For example, for
\[
\bar{S}_{1, n}(x_{1:n}, c_{1:n}^{d}, y_{1:n}) = \sum_{t = 1}^{n} c_{t}^{d} x_{t} x_{t}^{T},
\]
we perform forward smoothing for each
\[
\bar{S}_{1, n, ij}(x_{1:n}, c_{1:n}^{d}, y_{1:n}) = \sum_{t = 1}^{n} c_{t}^{d} x_{t}(i) x_{t}(j), \quad i, j = 1, \ldots, d_{x}.
\]
It was shown in \citet{Elliott_and_Krishnamurthy_1999} that, the intermediate function
\[
\bar{T}_{1, t, ij}^{\theta}(x_{t}, c_{1:t}^{d}) := \mathbb{E}_{\theta}\left[ \left. \bar{S}_{1, t, ij}(X_{1:t}, c_{1: t}^{d}, y_{1:t}) \right\vert c_{1: t}^{d}, x_t, y_{1:t}  \right]
\]
for the $i, j$'th element is a quadratic in $x_{t}$:
\begin{equation}
\bar{T}_{1, t, ij}^{\theta}(x_{t}, c_{1:t}^{d}) = x_{t}^{T} \bar{P}_{1, t, ij} x_{t} + \bar{q}_{1, t, ij}^{T} x_{t} + \bar{r}_{1, t, ij},
\end{equation}
where $\bar{P}_{1, t, ij}$ is a $d_{x} \times d_{x}$ matrix, $\bar{q}_{1, t, ij}$ is a $d_{x} \times 1$ vector, and $\bar{r}_{1, t, ij}$ is a scalar. Online smoothing is then performed via the following recursion over the variables $\bar{P}_{1, t, ij}, \bar{q}_{1, t, ij}, \bar{r}_{1, t, ij}$.
\begin{align*} \label{eq: forward smoothing recursion in GLSSM}
\bar{P}_{1, t+1, ij} & = B_{t}^{T} \bar{P}_{1, t, ij} B_{t} + c_{t+1}^{d} e_{i}e_{j}^{T}, \\
\bar{q}_{1, t+1, ij} & = B_{t}^{T} \bar{q}_{1, t, ij} +  B_{t}^{T} \left( \bar{P}_{1, t, ij} +  \bar{P}_{1, t, ij}^{T} \right) b_{t}, \\
\bar{r}_{1, t+1, ij} & = \bar{r}_{1, t, ij} + \text{tr} \left(\bar{P}_{1, t, ij} \Sigma_{t | t+1} \right) + \bar{q}_{1, t, ij}^{T} b_{t} +  b_{t}^{T} \bar{P}_{1, t,ij} b_{t},
\end{align*}
where $e_{i}$ is the $i$'th column of the identity matrix of the size $d_{x}$, and $\text{tr}(A)$ is the trace of the matrix $A$. For the initial value of $\bar{T}_{1, 1, ij}^{\theta}(x_{1}, c_{1}^{d})$, $\bar{P}_{1, 1, ij} = c_{1}^{d} e_{i}e_{j}^{T}, q_{1, 1, ij} = 0_{d_{x} \times 1}, \bar{r}_{1, 1, ij} = 0 $. Therefore, the $i, j$'th element of the required expectation at time $n$ can be calculated as
\begin{align*}
&\mathbb{E}_{\theta} \left[ \left. \bar{T}_{1, n, ij}^{\theta}(X_{n}, c_{1:n}^{d}) \right\vert y_{1:n}, c_{1:n}^{d} \right] = \\
& \quad\quad \text{tr} \left(\bar{P}_{1, n, ij} \left( \Sigma_{n|n} + \mu_{n|n} \mu_{n|n}^{T} \right) \right) + \bar{q}_{1, n, ij}^{T} \mu_{n|n} + \bar{r}_{1, n, ij}. \nonumber
\end{align*}
We can similarly obtain the recursions for the other sufficient statistics in terms of variables $\bar{P}_{m, t, ij}, \bar{q}_{m, t, ij}, \bar{r}_{m, t, ij}$ for the $m$'th sufficient statistic (see Appendix \ref{sec: Recursive updates for sufficient statistics in a single GLSSM}) \citep{Elliott_and_Krishnamurthy_1999}.

\begin{rem}
Note that $\bar{P}_{1, t, ji}=(\bar{P}_{1, t, ij})^T$ (similarly for $\bar{q}_{1, t, ij}$) and therefore need only be calculated for $j \geq i$.
Note that the variables $\mu_{t | t}, \Sigma_{t |t}, \Gamma_{t}, \epsilon_{t}, B_{t}, b_{t}, \Sigma_{t | t+1}, \bar{P}_{m, t, ij}, \bar{q}_{m, t, ij}, \bar{r}_{m, t, ij}$ obviously depend on $c_{1:t}^{d}$, $y_{1:t}$ and $\theta$, but we made this dependancy implicit in our notation for simplicity. We will carry on with this simplification in the rest of the paper.
\end{rem}

\subsubsection{Application to MTT} \label{sec: Application to MTT}
We showed above how to calculate expectations of the required sufficient for a single GLSSM. We can extend that idea to the scenario in the MTT case, where there may be multiple GLSSMs at a time, with different starting and ending times and possible missing observations. Recall that  at time $t$ the targets which are alive are the $k_{t}^{s}$ surviving targets from $t-1$ and the $k_{t}^{b}$ newly born targets at time $t$, so the number of targets is $k_{t}^{x} = k_{t}^{s} + k_{t}^{b}$. For each alive target, we can calculate the moments of the prediction density $p_{\theta}(x_{t, k} | \mathbf{y}_{1:t-1}, z_{1:t})$ for the state
\begin{align}
(\mu_{t | t-1, k}, \Sigma_{t | t-1, k}) \hspace{-0.1cm} = \hspace{-0.1cm}
\begin{cases}
	\begin{aligned}
		& \hspace{-0.2cm} \left( F \mu_{t-1 | t-1, i_{t}^{s}(k)}, \right. \\
		& \left. F  \Sigma_{t | t-1, i_{t}^{s}(k)} F^{T} + W \right) \\
	\end{aligned}, & \hspace{-0.3cm}  k \leq k_{t}^{s}, \\
\left( \mu_{b}, \Sigma_{b} \right), & \hspace{-0.4cm} k_{t}^{s} < k \leq k_{t}^{x}
\end{cases}. \nonumber
\end{align}
Recall that $i^{s}_{t}(k)$ appears above due to the relabelling of surviving targets from time $t-1$. Also, given the detection vector $c_{t}^{d}$ and the association vector $a_{t}$, we calculate the moments of the filtering density $p_{\theta}(x_{t, k} | \mathbf{y}_{1:t}, z_{1:t})$ for the targets using the prediction moments
\begin{align}
& (\mu_{t | t, k}, \Sigma_{t | t, k}) = \nonumber\\
& \begin{cases}
	\begin{aligned}
		& \left( \mu_{t | t-1, k} + \Sigma_{t | t-1, k} G^{T} \Gamma_{t, k}^{-1} \epsilon_{t, k},\right.\\
		& \quad\quad\left.\Sigma_{t | t-1, k}  - \Sigma_{t | t-1, k} G^{T} \Gamma_{t, k}^{-1} G \Sigma_{t | t-1, k}  \right) \\
	\end{aligned},  & c_{t}^{d}(k) = 1 \\
	\left( \mu_{t | t-1, k}, \Sigma_{t | t-1, k} \right), & c_{t}^{d}(k) = 0.
\end{cases} \nonumber
\end{align}
where $\Gamma_{t, k} = G \Sigma_{t | t-1, k} G^{T} + V$ and $\epsilon_{t, k} = y_{t, a_{t}(i'_{t}(k))} - G \mu_{t | t-1, k}$, where $i'_{t}(k) = \sum_{j = 1}^{k} c_{t}^{d}(j)$. Note that if the $k$'th  alive target at time $t$ is detected, it will be the $i'_{t}(k)$'th detected target, which explains $i'_{t}(k)$ in $\epsilon_{t, k}$. In a similar manner, we calculate $B_{t, k} $, $b_{t, k}$, and $\Sigma_{t | t+1, k}$ using $\mu_{t | t, k}$ and $\Sigma_{t | t, k}$ for $k = 1, \ldots, k_{t}^{x}$ in analogy with $B_{t}$, $b_{t}$, and $\Sigma_{t | t+1}$.

In the following, we will present the rules for one-step update of the expectations
\begin{equation*}
\widetilde{S}_{m, n}^{\theta}(z_{1:n}) = \mathbb{E}_{\theta} \left[ \left. S_{m, n}(\mathbf{X}_{1:n}, z_{1:n}) \right\vert \mathbf{y}_{1:n}, z_{1:n} \right]
\end{equation*}
of the sufficient statistics $S_{m, n}(\mathbf{x}_{1:n}, z_{1:n})$ that are defined in \eqref{eq: KF sufficient statistics for MTT}. Observe that we can write for $\quad 1 \leq m \leq 7$,
\begin{align}
S_{m, n}(\mathbf{x}_{1:n}, z_{1:n}) & =  s_{m}(\mathbf{x}_{1}, z_{1}) + \sum_{t = 2}^{n} s_{m}(\mathbf{x}_{t-1}, \mathbf{x}_{t}, z_{t} ),
\end{align}
where the functions $s_{m}$ can be written in terms of $\bar{s}_{m}$'s (\ref{eq:barsm}) as follows:
\begin{align}
& s_{m}(\mathbf{x}_{1}, z_{1}) =  \sum_{k = 1}^{k_{1}^{b}} \bar{s}_{m}(x_{1, k}, c_{1}^{d}(k), y_{1, a_{1}(i'_{1}(k))}), \nonumber \\
& s_{m}(\mathbf{x}_{t-1}, \mathbf{x}_{t}, z_{t}) = \sum_{k = 1}^{k_{t}^{s}}  \bar{s}_{m}(x_{t-1, i_{t}^{s}(k)}, x_{t, k}, c_{t}^{d}(k), y_{t, a_{t}(i'_{t}(k)) })\nonumber\\
&\quad\quad\quad\quad\quad\quad\quad\quad\quad\quad + \sum_{k = k_{t}^{s} + 1}^{k_{t}^{x}} \bar{s}_{m}(x_{t, k}, c_{t}^{d}(k), y_{t, a_{t}(i'_{t}(k))}). \nonumber
\end{align}
where, again, $i'_{t}(k) = \sum_{j = 1}^{k} c_{t}^{d}(j)$. (Notice that if $c_{t}^{d}(k) = 0$ this $i'_{t}(k)$ can still be used as a convention; since the choice of the observation point in $\mathbf{y}_{t}$ is irrelevant as it will have no contribution being multiplied by $c_{t}^{d}(k)$.) Therefore, the forward smoothing recursion for those sufficient statistics in \eqref{eq: KF sufficient statistics for MTT} at time $t$
\begin{equation}
\begin{split}
T_{m, t}^{\theta}(\mathbf{x}_{t}, z_{1:t}) &= \mathbb{E}_{\theta} \left[T_{m, t-1}^{\theta}(\mathbf{X}_{t-1}, z_{1:t-1})\right.\nonumber\\
&\quad\quad \quad \left.+s_{m}\left( \mathbf{X}_{t-1}, \mathbf{x}_{t}, z_{t} \right)\vert \mathbf{x}_{t}, \mathbf{y}_{1:t-1}, z_{1:t-1} \right] \label{eq: recursive update of T in MTT}
\end{split}
\end{equation}
can be handled once we have the forward smoothing recursion rules for the sufficient statistics in \eqref{eq: KF sufficient statistics}. For $k = 1, \ldots, k_{t}^{x}$, let $T_{m, t, k}^{\theta}$ denote the forward smoothing recursion function for the $m$'th sufficient statistic for $k$'th alive target at time $t$. For the surviving targets, $k$'th target at time $t$ is a continuation of the $i_{t}^{s}(k)$'the target at time $t-1$. Therefore, we have the recursion update for $T_{m, t, k}^{\theta}$ for $1 \leq k \leq k_{t}^{s}$ as
\begin{equation*}
\begin{split}
& T_{m, t, k}^{\theta}(x_{t, k}, z_{1:t}) = \mathbb{E}_{\theta} \left[ T_{m, t-1, i_{t}^{s}(k)}^{\theta}(X_{t-1, i_{t}^{s}(k)}, z_{1:t-1}) \right. \\
& \quad \left. + \bar{s}_{m}(X_{t-1, i_{t}^{s}(k)}, x_{t, k}, c_{t}^{d}(k), y_{a_{t}(i'_{t}(k))}) \right\vert  x_{t, k}, \mathbf{y}_{1:t-1}, z_{1:t-1} \Big].
\end{split}
\end{equation*}
For the targets born at time $t$ (for $k_{t}^{s} + 1 \leq k \leq k_{t}^{x}$ ), the recursion function is initiated as $T_{m, t, k}^{\theta}(x_{t, k}, z_{1:t}) = s_{m}(x_{t, k}, c_{t}^{d}(k))$. Therefore, the $(i, j)$'th component of the recursion function can be written as
\[
T_{m, t, k, ij}^{\theta}(x_{t, k}, z_{1:t}) = x_{t, k}^{T}P_{m, t, k, ij} x_{t, k} + q_{m, t, k, ij}^{T} x_{t, k} + r_{m, t, k, ij}
\]
similarly to the single GLSSM case, where this time we have the additional subscript $k$. For surviving targets the recursion variables $P_{m, t, k, ij}, q_{m, t, k, ij}, r_{m, t, k, ij}$ for each $m, i, j$ are updated from $P_{m, t-1, i_{t}^{s}(k), ij}, q_{m, t-1, i_{t}^{s}(k), ij}, r_{m, t-1, i_{t}^{s}(k), ij}$, by using $\mu_{t-1 | t-1, i_{t}^{s}(k)}$, $\Sigma_{t-1 | t-1, i_{t}^{s}(k)}$, $B_{t-1, i_{t}^{s}(k)}$, $b_{t-1, i_{t}^{s}(k)}$, $\Sigma_{t-1 | t, i_{t}^{s}(k)}$, $c_{t}^{d}(k)$ and, $y_{t, a_{t}(i'_{t}(k))}$ with $ i'_{t}(k) = \sum_{j = 1}^{k} c_{t}^{d}(j)$. For the targets born at time $t$ (for $k_{t}^{s} + 1 \leq k \leq k_{t}^{x}$ ), the variables are set to their initial values in the same way as in Section \ref{sec: Online smoothing in a single GLSSM} using $c_{t}^{d}(k)$ and, if $c_{t}^{d}(k) = 1$, $y_{t, a_{t}(i'_{t}(k))}$. The conditional expectations of sufficient statistics
\[
\widetilde{S}_{m, t}^{\theta}(z_{1:t}) = \mathbb{E}_{\theta} \left[ \left. T_{m, t}^{\theta} \left( \mathbf{X}_{t}, z_{1:t}\right) \right\vert \mathbf{y}_{1:t}, z_{1:t} \right]
\]
 can then be calculated by using the forward recursion variables and the filtering moments. Let
\[
\widetilde{S}_{m, t, k}^{\theta}(z_{1:t}) = \mathbb{E}_{\theta} \left[ \left. T_{m, t, k}^{\theta}(X_{t, k}, z_{1:t}) \right\vert \mathbf{y}_{1:t}, z_{1:t} \right]
\]
denote the expectation of the $m$'th sufficient statistic for the $k$'th alive target at time $t$, where its $(i, j)$'th component is
\begin{equation*}
\begin{split}
\widetilde{S}_{m, t, k, ij}^{\theta}(z_{1:t}) &= \text{tr} \left(P_{m, t, k, ij} \left(  \mu_{t | t, k} \mu_{t | t, k}^{T} + \Sigma_{t |t, k} \right) \right) \\
&\quad\quad\quad\quad\quad\quad\quad + q_{m, t, k, ij}^{T} \mu_{t |t, k} + r_{m, t, k, ij}.
\end{split}
\end{equation*}
Then, the required conditional expectation for the $m$'th sufficient statistic can be written as the sum of two quantities
\begin{align}
\widetilde{S}_{m, t}^{\theta}(z_{1:t}) &= \widetilde{S}_{alive, m, t}^{\theta}(z_{1:t}) + \widetilde{S}_{dead, m, t}^{\theta}(z_{1:t}). \label{eq: expectation as sum of contributions of alive and dead targets}
\end{align}
where the quantities are respectively the contributions of the alive targets at time $t$ and dead targets up to time $t$ to the conditional expectation $\widetilde{S}_{m, t}^{\theta}(z_{1:t})$
\begin{align} \label{eq: contribution of alive and dead targets}
\widetilde{S}_{alive, m, t}^{\theta}(z_{1:t}) &= \sum_{k = 1}^{k_{t}^{x}} \widetilde{S}_{m, t, k}^{\theta}(z_{1:t}), \nonumber\\
\widetilde{S}_{dead, m, t}^{\theta}(z_{1:t}) &= \sum_{j = 1}^{t} \sum_{k: c_{j}^{s}(k) = 0}  \widetilde{S}_{m, j-1, k}^{\theta}(z_{1:j-1})
\end{align}
As \eqref{eq: expectation as sum of contributions of alive and dead targets} shows, we also need to calculate $\widetilde{S}_{dead, m, t}^{\theta}(z_{1:t})$ at each time and by \eqref{eq: contribution of alive and dead targets} this can easily be done by storing $\widetilde{S}_{dead, m, t-1}^{\theta}(z_{1:t-1})$ at time $t-1$ and using the recursion
\[
\widetilde{S}_{dead, m, t}^{\theta}(z_{1:t}) \hspace{-0.1cm} = \hspace{-0.1cm} \widetilde{S}_{dead, m, t-1}^{\theta}(z_{1:t-1}) + \hspace{-0.3cm} \sum_{k: c_{t}^{s}(k) = 0}  \widetilde{S}_{m, t-1, k}^{\theta}(z_{1:t-1})
\]
\normalsize
where the terms in the sum correspond to targets that terminate at time $t-1$.

Finally, the sufficient statistics $S_{8, n}(z_{1:n}), \ldots, S_{15, n}(z_{1:n})$ can be calculated online since we can write for each $m = 8, \ldots, 15$
\[
S_{m, n}(z_{1:n}) = \sum_{t = 1}^{n} s_{m}(z_{t})
\]
for some suitable functions $s_{m}$ which can easily be constructed from \eqref{eq: MTT sufficient statistics}. Hence they can be updated online as
\begin{equation} \label{eq: MTT sufficient statistics update}
S_{m, t}(z_{1:t}) = S_{m, t-1}(z_{1:t-1}) + s_{m}(z_{t}).
\end{equation}

We now present Algorithm \ref{alg: One step update for sufficient statistics in the MTT model} to show how these one-step update rules for the sufficient statistics in the MTT model can be implemented. For simplicity of the presentation, we will use a short hand notation for representing the forward recursion variables in a batch way. Let $\mathcal{T}_{m, t}^{\theta}(z_{1:t}) = ( \mathcal{T}_{m, t, k}^{\theta}(z_{1:t}), k = 1, \ldots, k^{x}_{t} )$ where
\[
\mathcal{T}_{m, t, k}^{\theta}(z_{1:t}) = \left( P_{m, t, k, ij}, q_{m, t, k, ij}, r_{m, t, k, ij}: \text{all } i, j \right)
\]
denote all the variables required for the forward smoothing recursion for the $m$'th sufficient statistic for the $k$'th alive target at time $t$. We can now present the algorithm using this notation.
\small
\begin{alg} \label{alg: One step update for sufficient statistics in the MTT model}
\textbf{One step update for sufficient statistics in the MTT model} \\
We have $ \mathcal{T}_{m, t-1}^{\theta}(z_{1:t-1})$, $\widetilde{S}_{dead, m, t-1}^{\theta}(z_{1:t-1})$,  $m =1, \ldots, 7$,  $S_{m^{\prime}, t-1}^{\theta}(z_{1:t-1})$, $m^{\prime} = 8, \ldots, 15$ at time $t-1$.
Given $z_{t}$ and $\mathbf{y}_{t}$, \\
- Set $i_{x} = 0$, $i_{d} = 0$, $\widetilde{S}_{alive, m, t}^{\theta}(z_{1:t}) = 0$ and $\mathcal{S}_{dead, m, t}^{\theta}(z_{1:t}) = \mathcal{S}_{dead, m, t-1}^{\theta}(z_{1:t-1})$ for $m = 1, \ldots, 7$. \\
- for $i = 1, \ldots, k^{x}_{t-1} + k^{b}_{t}$
\begin{itemize}
\item if $i \leq k^{x}_{t-1}$ and $c^{s}_{t}(i) = 1$, (the $i$'th target at time $t-1$ survives), or if $i > k^{x}_{t-1}$, (a new target is born), set $i_{x} = i_{x} + 1$.
\begin{itemize}
\item In case of survival, use $\mu_{t-1 | t-1, i}$ and $\Sigma_{t-1 | t-1, i}$ to obtain the prediction moments $\mu_{t | t-1, i_{x}}$ and $\Sigma_{t | t-1, i_{x}}$. In case of birth, set the prediction distribution $\mu_{t | t-1, i_{x}} = \mu_{b}$ and $\Sigma_{t|t-1, i} = \Sigma_{b}$.
\begin{itemize}
\item If $c^{d}_{t}(i_{x}) = 1$, $i_{x}$'th target is detected: $i_{d} = i_{d} + 1$. Use $\mu_{t | t-1, i_{x}}$ and $\Sigma_{t | t-1, i_{x}}$ and $y_{t, a_{t}( i_{d})}$ to update the filtering moments $\mu_{t | t, i_{x}}$ and $\Sigma_{t |t, i_{x}}$.
\item If $c^{d}_{t}(i_{x}) = 0$, $i_{x}$'th target is not detected: Set $\left( \mu_{t | t, i_{x}}, \Sigma_{t|t, i_{x}} \right) = \left( \mu_{t | t-1, i_{x}}, \Sigma_{t|t-1, i_{x}} \right)$.
\end{itemize}
\item For $m = 1, \ldots, 7$
\begin{itemize}
\item In case of survival, update the recursion variables $\mathcal{T}_{m, t, i_{x}}^{\theta}(z_{1:t})$ using $\mathcal{T}_{m, t-1, i}^{\theta}(z_{1:t-1})$, $\mu_{t-1 | t-1, i}$, $\Sigma_{t-1 | t-1, i}$, $b_{t-1, i}$, $B_{t-1, i}$, $\Sigma_{t-1 | t, i}$, $c_{t}^{d}(i_{x})$ and $y_{t, a_{t}(i_{d})}$ if $c_{t}^{d}(i_{x}) = 1$. In case of birth, initiate $\mathcal{T}_{m, t, i_{x}}^{\theta}(z_{1:t})$ using $c_{t}^{d}(i_{x})$ and $y_{t, a_{t}(i_{d})}$ if $c_{t}^{d}(i_{x}) = 1$.
\item \textbf{\emph{(optional)}} Calculate $\widetilde{S}_{m, t, i_{x}}^{\theta}(z_{1:t})$ using $\mathcal{T}_{m, t, i_{x}}^{\theta}(z_{1:t})$, $\mu_{t | t, i_{x}}$ and $\Sigma_{t | t, i_{x}}$ and update $\widetilde{S}_{alive, m, t}^{\theta}(z_{1:t}) \leftarrow \widetilde{S}_{alive, m, t}^{\theta}(z_{1:t}) +  \widetilde{S}_{m, t, i_{x}}^{\theta}(z_{1:t})$.
\end{itemize}
\end{itemize}
\item if $i \leq k^{x}_{t-1}$ and  $c^{s}_{t}(i) = 0$, the $i$'th target at time $t-1$ is dead. For $m = 1, \ldots, 7$,
\begin{itemize}
\item Calculate $\widetilde{S}_{m, t-1, i}^{\theta}(z_{1:t-1})$ from $\mathcal{T}_{m, t-1, i}(z_{1:t-1})$, $\mu_{t-1 | t-1, i}$ and $\Sigma_{t-1 | t-1, i}$.
\item Update $\widetilde{S}_{dead, m, t}^{\theta}(z_{1:t}) \leftarrow \widetilde{S}_{dead, m, t}^{\theta}(z_{1:t}) + \widetilde{S}_{m, t-1, i}^{\theta}(z_{1:t-1}).$
\end{itemize}
\end{itemize}
- \textbf{\emph{(optional)}} Update $ \widetilde{S}_{m, t}^{\theta}(z_{1:t}) = \widetilde{S}_{alive, m, t}^{\theta}(z_{1:t}) + \widetilde{S}_{dead, m, t}^{\theta}(z_{1:t}) $ for $m = 1, \ldots, 7$. \\
- Update $S_{m, t}(z_{1:t})  = S_{m, t-1}(z_{1:t-1}) + s_{m}(z_{t})$ for $m = 8, \ldots, 15$.
\end{alg}
\normalsize
Notice that the lines of the algorithm labeled as ``optional" are not necessary for the recursion and need not to be performed at every time step. For example, we can use Algorithm \ref{alg: One step update for sufficient statistics in the MTT model} in a batch EM to save memory, in that case we perform these steps only at the last time step $n$ to obtain the required expectations. Notice also that we included the update rule for the sufficient statistics in \eqref{eq: MTT sufficient statistics} for completeness.

\subsubsection{Online EM implementation} \label{sec: Online EM implementation}

In order to develop an online EM algorithm, we exploit the availability of calculating $\widetilde{S}_{1, t}^{\theta}, \ldots, \widetilde{S}_{7, t}^{\theta}$ and $S_{8, t}, \ldots, S_{15, t}$ in an online manner as shown in Section \ref{sec: Application to MTT}. In online EM, running averages of sufficient statistics are calculated and then used to update the estimate of $\theta^{\ast}$ at each time \citep{Elliott_et_al_2002, Mongillo_and_Deneve_2008, Cappe_2009a, Cappe_2011}. Let $\theta_{1}$ be the initial guess of $\theta^{\ast}$ before having made any observations and at time $t$, let $\theta_{1:t}$ be the sequence of parameter estimates of the online EM algorithm computed sequentially based on $\mathbf{y}_{1:t-1}$. When $\mathbf{y}_{t}$ is received, we first update the posterior density to have $\widehat{p}_{\theta_{1:t}}(z_{1:t} | \mathbf{y}_{1:t})$, and compute for $1 \leq m \leq 7$
\begin{align}
& T_{\gamma, m, t}^{\theta_{1:t}} \left( \mathbf{x}_{t}, z_{1:t} \right) = \mathbb{E}_{\theta_{1:t}} \Big[ (1 - \gamma_{t}) T_{\gamma, m, t-1}^{\theta_{1:t-1}} \left( \mathbf{X}_{t-1}, z_{1:t-1} \right) \nonumber\\
& \quad\quad\quad\quad\quad + \gamma_{t} s_{m}\left( \mathbf{X}_{t-1}, \mathbf{x}_{t}, z_{t} \right)  \Big\vert \mathbf{x}_{t}, \mathbf{y}_{1:t-1}, z_{1:t-1} \Big] \label{eq: stochastic approximation of forward smoothing}
\end{align}
for the values $z_{1:t} = z_{1:t}^{(i)}$ for $i = 1, \ldots, N$, where we have the same constraints on the step-size sequence $\left\{ \gamma_{t} \right\}_{t \geq 1}$ as in the SAEM algorithm. This modification reflects on the updates rules for the variables in $\mathcal{T}_{m, t}^{\theta}$. To illustrate the change in the recursions with an example, the recursion rules for the variables for $S_{1, t}(x_{1:t}, c_{1:t}^{d})$ for the simple GLSSM case become (see Appendix \ref{sec: Recursive updates for sufficient statistics in a single GLSSM})
\begin{align}
\bar{P}_{\gamma, 1, t+1, ij} & = (1 - \gamma_{t+1}) B_{t}^{T} \bar{P}_{\gamma,1, t, ij} B_{t} + \gamma_{t+1} c_{t+1}^{d} e_{i}e_{j}^{T} \nonumber \\
\bar{q}_{\gamma, 1, t+1, ij} & = (1 - \gamma_{t+1}) \Big( B_{t}^{T} \bar{q}_{\gamma, 1, t, ij}  \nonumber \\
& \quad\quad\quad\quad\quad\quad\quad\quad + B_{t}^{T} \left( \bar{P}_{\gamma, 1, t, ij} +  \bar{P}_{\gamma, 1, t, ij}^{T} \right) b_{t} \Big) \nonumber \\
\bar{r}_{\gamma, 1, t+1, ij} & = (1 - \gamma_{t+1})  \Big( \bar{r}_{\gamma, 1, t, ij} +  \text{tr} \left(\bar{P}_{\gamma, 1, t, ij} \Sigma_{t | t+1} \right) \nonumber \\
& \quad\quad\quad\quad\quad\quad\quad\quad\quad +  \bar{q}_{\gamma, 1, t, ij}^{T} b_{t} +  b_{t}^{T} \bar{P}_{\gamma, 1, t,ij} b_{t} \Big) \nonumber
\end{align}
\normalsize
So this time we have $\mathcal{T}_{\gamma, m, t}^{\theta_{1:t}}(z_{1:t}) = ( \mathcal{T}_{\gamma, m, t, k}^{\theta_{1:t}}(z_{1:t}), k = 1, \ldots, k_{t}^{x} )$ where
\[
\mathcal{T}_{\gamma, m, t, k}^{\theta_{1:t}}(z_{1:t}) = \left( P_{\gamma, m, t,k,  ij}, q_{\gamma, m, t,k,  ij}, r_{\gamma, m, t,k,  ij}: \text{all } i,j   \right).
\]
and the conditional expectations
\[
\widetilde{S}_{\gamma, m, t}^{\theta_{1:t}}(z_{1:t}) = \widetilde{S}_{\gamma, alive, m, t}^{\theta_{1:t}}(z_{1:t}) + \widetilde{S}_{\gamma, dead, m, t}^{\theta_{1:t}}(z_{1:t})
\]
can be calculated by using $\mathcal{T}_{\gamma, m, t, k}^{\theta_{1:t}}(z_{1:t})$ as in Section \ref{sec: Application to MTT}. Finally, regarding those $S_{m, t}$ in \eqref{eq: MTT sufficient statistics}, we calculate $8 \leq m \leq 15.$
\begin{align}
S_{\gamma, m, t} \left( z_{1:t} \right) = (1- \gamma_{t}) S_{\gamma, m, t-1} \left( z_{1:t-1} \right) + \gamma_{t} s_{m} \left( z_{t} \right). \label{eq: stochastic approximation for MTT sufficient statistics}
\end{align}
for the values $z_{1:t} = z_{1:t}^{(i)}$ for $i = 1, \ldots, N$.
In the maximisation step, we update $\theta_{t+1} = \Lambda \left( \widehat{S}_{\gamma, 1, t}^{\theta_{1:t}}, \ldots, \widehat{S}_{\gamma, 15, t}^{\theta_{1:t}} \right)$ where the expectations are obtained
\begin{align*}
\widehat{S}_{\gamma, m, t}^{\theta_{1:t}} = \begin{cases}
      \sum_{i = 1}^{N} w_{t}^{(i)}  \widetilde{S}_{\gamma, m, t}^{\theta_{1:t}} (z_{1:t}^{(i)} ), & 1 \leq m \leq 7, \\
       \sum_{i = 1}^{N} w_{t}^{(i)} S_{\gamma, m, t} ( z_{1:t}^{(i)} ), & 8 \leq m \leq 15.
\end{cases}
\end{align*}
In practice, the maximisation step is not executed until a burn-in time $t_{b}$ for added stability of the estimators (e.g. see \citet{Cappe_2009a}).

Notice that the SMC online EM algorithm can be implemented with the help of Algorithm \ref{alg: One step update for sufficient statistics in the MTT model} the only changes are \eqref{eq: stochastic approximation of forward smoothing} and \eqref{eq: stochastic approximation for MTT sufficient statistics} instead of \eqref{eq: recursive update of T in MTT} and \eqref{eq: MTT sufficient statistics update}. Algorithm \ref{alg: Online EM for the MTT model} describes the SMC online EM algorithm for the MTT model.
\small
\begin{alg} \label{alg: Online EM for the MTT model}
\textbf{The SMC online EM algorithm for the MTT model}
\begin{itemize}
\item \textbf{E-step:}
If $t = 1$, start with $\theta_{1}$, obtain $\widehat{p}_{\theta_{1}}(z_{1} | \mathbf{y}_{1}) = \sum_{i = 1}^{N} w_{1}^{(i)} \delta_{z_{1}^{(i)}} (z_{1})$, and for $i = 1, \ldots, N$ initialise \\
$\mathcal{T}_{\gamma, m, 1}^{\theta_{1}}( z_{1}^{(i)} )$, $\widetilde{S}_{\gamma, dead, m, 1}^{\theta_{1}}(z_{1}^{(i)} )$ for $m = 1, \ldots, 7$ and $S_{\gamma, m', 1} ( z_{1}^{(i)} )$ for $m' = 8, \ldots, 15 $,

If $t \geq 1$, \\
Obtain $\widehat{p}_{\theta_{1:t}}(z_{1:t} | \mathbf{y}_{1:t}) = \sum_{i = 1}^{N} w_{t}^{(i)} \delta_{z_{1:t}^{(i)}} (z_{1:t})$ from $\widehat{p}_{\theta_{1:t-1}}(z_{1:t-1} | \mathbf{y}_{1:t-1})$ along with $\pi_{t}$. \\
For $i = 1, \ldots, N$, set $j = \pi_{t}(i)$. Use Algorithm \ref{alg: One step update for sufficient statistics in the MTT model} with the stochastic approximation to obtain \\
$ \mathcal{T}_{\gamma, m, t}^{\theta_{1:t}}( z_{1:t}^{(i)} )$, $\widetilde{S}_{\gamma, dead, m, t}^{\theta_{1:t}}(z_{1:t}^{(i)} )$ for $m = 1, \ldots, 7$ and $S_{\gamma, m', t} ( z_{1:t}^{(i)} )$ for $m' = 8, \ldots, 15 $ from \\
 $\mathcal{T}_{\gamma, m, t-1}^{\theta_{1:t-1}}( z_{1:t-1}^{(j)} )$, $\widetilde{S}_{\gamma, dead, m, t-1}^{\theta_{1:t-1}}(z_{1:t-1}^{(j)} )$ for $m = 1, \ldots, 7$ and $S_{\gamma, m', t-1} ( z_{1:t-1}^{(j)} )$ for $m' = 8, \ldots, 15$.
\item \textbf{M-step:} If $t < t_{b}$, $\theta_{t+1} = \theta_{t}$. Else, for $i = 1, \ldots, N$, $m = 1, \ldots, 7$ calculate $ \widetilde{S}_{\gamma, alive, m, t}^{\theta_{1:t}}(z_{1:t}^{(i)} )$ and $\widetilde{S}_{\gamma, m, t}^{\theta_{1:t}} (z_{1:t}^{(i)} ) = \widetilde{S}_{\gamma, alive,  m, t}^{\theta_{1:t}} (z_{1:t}^{(i)} ) + \widetilde{S}_{\gamma, dead, m, t}^{\theta_{1:t}} (z_{1:t}^{(i)} ) $  (`\textbf{optional}' lines in Algorithm \ref{alg: One step update for sufficient statistics in the MTT model}). Calculate the expectations
\begin{equation*}
\begin{split}
&\left[ \widehat{S}_{\gamma, 1, t}^{\theta_{1:t}}, \ldots, \widehat{S}_{\gamma, 15, t}^{\theta_{1:t}} \right] \\
& \quad = \sum_{i = 1}^{N} w_{n}^{(i)} \left[ \widetilde{S}_{\gamma, m, t}^{\theta}, \ldots, \widetilde{S}_{\gamma, 7, t}^{\theta_{1:t}},  S_{\gamma, 8, t}, \ldots,  S_{\gamma, 15, t}  \right] \left( z_{1:t}^{(i)} \right).
\end{split}
\end{equation*}
\normalsize
and update $\theta_{t+1} = \Lambda \left( \widehat{S}_{\gamma, 1, t}^{\theta_{1:t}}, \ldots, \widehat{S}_{\gamma, 15, t}^{\theta_{1:t}} \right)$.
\end{itemize}
\end{alg}
\normalsize
Finally, before ending this section, we list in Table \ref{table: EM variables} some important variables used to describe the EM algorithms throughout the section.
\begin{table}
\caption{The list of the EM variables used in Section \ref{sec: EM algorithms for MTT} }
\vspace{0.1cm}
\begin{tabular}{| l |}
\hline
\textbf{Sections \ref{sec: Batch EM for MTT} and \ref{sec: Estimation of sufficient statistics}} \\
$S_{m, n}$, $m = 1:15$, Sufficient statistics of the MTT model \\
$S_{m, n}^{\theta}$, $m = 1:15$,  Expectation of $S_{m, n}$ conditional to $\mathbf{y}_{1:n}$ \\
$\widetilde{S}_{m, n}^{\theta}$, $m = 1:7$, Expectation of $S_{m, n}$ conditional to $\mathbf{y}_{1:n}$ and $z_{1:n}$ \\
\textbf{Section \ref{sec: Stochastic versions of EM}} \\
$\widehat{S}_{m, n}^{\theta}$, Monte Carlo estimation of  $S_{m, n}^{\theta}$ \\
$\widehat{S}_{\gamma, m, n}^{(j)}$, Weighted average of $\widehat{S}_{m, n}^{\theta_{1}}, \ldots,  \widehat{S}_{m, n}^{\theta_{j}}$ for the SAEM algorithm \\
\textbf{Section \ref{sec: Online smoothing in a single GLSSM}} \\
$\bar{S}_{m, n}$, $m = 1:7$,  Sufficient statistics of a single GLSSM \\
$\bar{s}_{m, t}$, $m = 1:7$, Incremental functions for $\bar{S}_{m, n}$ \\
$\bar{S}_{m, n, ij}$, The $(i, j)$'th element of $\bar{S}_{m, n}$ \\
$\bar{s}_{m, t, ij}$, The $(i, j)$'th element of $\bar{s}_{m, t}$ \\
$\bar{T}_{m, t, ij}$, Forward smoothing recursion (FSR) function for $\bar{S}_{m, t, ij}$ \\
$\bar{P}_{m, t, ij}, \bar{q}_{m, t, ij}, \bar{r}_{m, t, ij}$, Variables used to write $\bar{T}_{m, t, ij}$ in closed-form \\
\textbf{Section \ref{sec: Application to MTT}} \\
$s_{m, t}$, $m = 1:15$, Incremental functions for $S_{m, n}$ \\
$T_{m, t}^{\theta}$, $m = 1:7$, FSR function for $S_{m, t}$ \\
$T_{m, t, k}^{\theta}$, FSR function for $m$'th sufficient statistic of the $k$'th alive target\\
\hspace{0.95cm} at time $t$ \\
$T_{m, t, k, ij}^{\theta}$, The $(i, j)$th element of $T_{m, t, k}^{\theta}$ \\
$P_{m, t, k, ij}, q_{m, t, k, ij}, r_{m, t, k, ij}$, Variables to write $T_{m, t, k, ij}$ \\
$\widetilde{S}_{m, t, k}^{\theta}$ Expectation of the $m$'th sufficient statistic of the $k$'th alive target\\ \hspace{1cm} at time $t$\\
$\widetilde{S}_{m, t, k, ij}^{\theta}$, The $(i, j)$'th element of  $\widetilde{S}_{m, t, k}^{\theta}$ \\
$\widetilde{S}_{alive, m, t}^{\theta}$, Contributions of the alive targets at time $t$ to $\widetilde{S}_{m, t}^{\theta}$\\
$\widetilde{S}_{dead, m, t}^{\theta}$, Contributions of the dead targets up to time $t$ to $\widetilde{S}_{m, t}^{\theta}$\\
\textbf{Section \ref{sec: Online EM implementation}} \\
$T_{\gamma, m, t}^{\theta_{1:t}}$, Online estimation of $T_{m, t}^{\theta}$ using $\theta_{1:t}$ \\
$P_{\gamma, m, t, k, ij}, q_{\gamma, m, t, k, ij}, r_{\gamma, m, t, k, ij}$: Variables to write $T_{\gamma, m, t, k, ij}$ \\
$\widetilde{S}_{\gamma, alive, m, t}^{\theta_{1:t}}$, Online estimation of $\widetilde{S}_{alive, m, t}^{\theta}$ using $\theta_{1:t}$ \\
$\widetilde{S}_{\gamma, dead, m, t}^{\theta_{1:t}}$, Online estimation of $\widetilde{S}_{dead, m, t}^{\theta}$ using $\theta_{1:t}$ \\
$\widetilde{S}_{\gamma, m, t}^{\theta_{1:t}}$, Online estimation of $\widetilde{S}_{m, t}^{\theta}$ using $\theta_{1:t}$ \\
$ S_{\gamma, m, t}$, $m = 8:15$, Online calculation of $S_{m, n}$ using $\theta_{1:t}$ \\
$\widehat{S}_{\gamma, m, t}^{\theta_{1:t}}$, Online estimation of $\widehat{S}_{m, t}^{\theta}$ using $\theta_{1:t}$ \\
\hline
\end{tabular}
\label{table: EM variables}
\end{table}

\section{Experiments and results} \label{sec: Experiments and results}
We compare the performance of the parameter estimation methods described in Section \ref{sec: EM algorithms for MTT} for the constant velocity model in Example \ref{ex: The constant velocity model}, where the parameter vector is
\[
\theta = \left( \lambda_{b}, \lambda_{f}, p_{d}, p_{s}, \mu_{bp}, \mu_{bv}, \sigma_{bp}^{2}, \sigma_{bv}^{2}, \sigma_{xp}^{2}, \sigma_{xv}^{2}, \sigma_{y}^{2} \right).
\]
Note that the constant velocity model assumes the position noise variance $\sigma_{xp}^{2} = 0$. All other parameters are estimated.

\subsection{Batch setting} \label{sec: Batch setting}
\subsubsection{Comparison of methods for batch estimation} \label{sec: Comparison of methods for batch estimation}
We run two experiments using the constant velocity model in the batch setting. In the first experiment, we generate an observation sequence of length $n = 100$ by using the parameter value
\[
\theta^{\ast} = (0.2, 10, 0.90, 0.95, 0, 0, 25, 4, 0, 0.0625, 4)
\]
and window size $\kappa = 100$. This particular value of $\theta^{\ast}$ creates on average $1$ target every $5$ time steps, and the average life of a target is $20$ time steps. Therefore we expect to see around $4$ targets per time.

Using the generated data set, we compare the performance of the three different methods for batch estimation, which are SMC-EM and MCMC-EM (two different implementations of SAEM in Algorithm \ref{alg: SAEM for the MTT model}) for MLE, and MCMC for the Bayesian estimation \citep{Yoon_and_Singh_2008}. For SMC-EM, we used $N = 200$ particles to implement the SMC method based on the $L$-best linear assignment to sample associations, where we set $L = 10$, the details of the SMC method are in Appendix \ref{sec: SMC algorithm for MTT}. For the MCMC-EM, in each EM iteration we ran $5$ MCMC steps and the last sample is taken to compute the sufficient statistics, i.e.\ $N = 1$. For both the SMC and MCMC implementations of SAEM, $\gamma_{j} = j^{-0.8}$ is used as the sequence of step-sizes for all parameters to be estimated, with the exception that $\gamma_{j} = j^{-0.55}$ is used for estimating $\sigma_{xv}^{2}$. That is to say, in the SAEM algorithm, $\widehat{S}_{\gamma, 3, n}^{(j)}$, $\widehat{S}_{\gamma, 4, n}^{(j)}$, and $\widehat{S}_{\gamma, 5, n}^{(j)}$ are calculated using $\gamma_{j} = j^{-0.55}$, and $\widehat{S}_{\gamma, 11, n}^{(j)}$ is calculated twice by using $\gamma_{j} = j^{-0.55}$ and $\gamma_{j} = j^{-0.8}$ separately (since it appears both in the estimation of $\sigma_{xv}^{2}$ and $p_{s}$), and for the rest of $\widehat{S}_{\gamma, m, n}^{(j)}$ $\gamma_{j} = j^{-0.8}$ is used. For Bayesian estimation, the following conjugate priors are used:
\begin{align*}
&p_{s}, p_{d} \stackrel{iid}{\sim} \text{Unif}\,(0,1), \quad \lambda_{b}, \lambda_{f} \stackrel{iid}{\sim}  \mathcal{G}(0.001, 1000), \\
&\sigma_{xv}^{2},  \sigma_{y}^{2},  \sigma_{bp}^2, \sigma_{bv}^{2} \stackrel{iid}{\sim} \mathcal{IG}(0.001, 0.001), \\
&\mu_{bx} | \sigma_{bp}^{2} \sim \mathcal{N}(0.1, 1000 \sigma_{bp}^2), \quad \mu_{by} | \sigma_{bp}^{2} \sim \mathcal{N}(-0.1, 1000 \sigma_{bp}^{2} ).
\end{align*}
\begin{figure*}[t!]
\centerline{\epsfig{figure=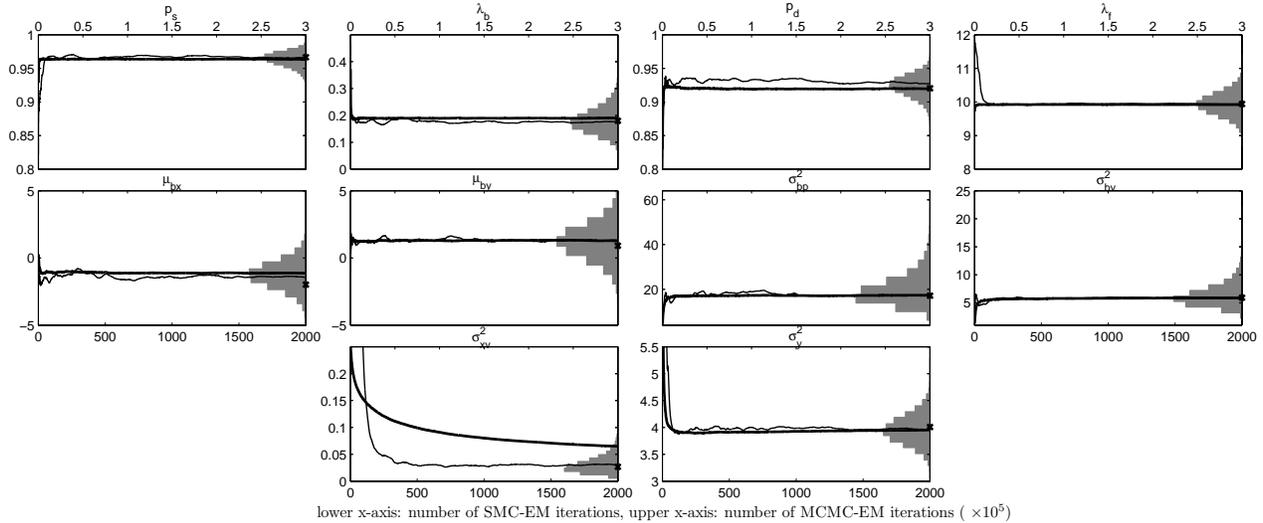, scale = 0.50}}
\caption{Batch estimates obtained using the SMC-EM (thin lines) and MCMC-EM (bold lines) algorithms for MLE and MCMC algorithm for the Bayesian estimate (histograms). $\theta^{\ast, z}$ is shown as a cross. Upper and lower x-axes show the number of EM iterations for MCMC-EM and SMC-EM, respectively.}
\label{fig: SAEM_estimates_for_MTT}
\end{figure*}
Figure \ref{fig: SAEM_estimates_for_MTT} shows the results obtained using SMC-EM, MCMC-EM and MCMC after $2000$, $3\times 10^5$, $3\times 10^5$ iterations respectively. For the Bayesian estimate, we consider only the last $5000$ samples generated using  MCMC as samples from the true posterior $p(\theta | \mathbf{y}_{1:n})$. For comparison, we also execute the EM algorithm with the true data association and the resulting $\theta^{\ast}$ estimate will serve as the benchmark. Note that given the true association, the EM can be executed without the need for any Monte Carlo approximation, and it gave the estimate
\begin{equation*}
\begin{split}
& \theta^{\ast, z} =  (0.18, 9.94, 0.92, 0.97, -1.98, 0.91, 17.18, 5.92, \\
& \quad\quad\quad\quad\quad\quad\quad\quad\quad\quad\quad\quad\quad\quad\quad\quad\quad\quad  0, 0.027, 4.01).
\end{split}
\end{equation*}
The $z$ in the superscript is to indicate that this value of $\theta$ maximises the joint probability density of $\mathbf{y}_{1:n}$ and $z_{1:n}$, i.e.
\[
\theta^{\ast, z} = \arg \max_{\theta \in \Theta} \log p_{\theta}(\mathbf{y}_{1:n}, z_{1:n})
\]
which is different than $\theta_{\text{ML}}$. However, for a data size of $100$, $\theta^{\ast, z}$ is expected to be closer to $\theta_{\text{ML}}$ than $\theta^{\ast}$ is, hence it is useful for evaluating the performances of the stochastic EM algorithms we present. From Figure \ref{fig: SAEM_estimates_for_MTT}, we can see that almost all MLE estimates obtained using SMC-EM and MCMC-EM converge to values around $\theta^{\ast, z}$, except for $\sigma_{xv}^2$  from SMC-EM has not converged within the experiment running time. The histogram of the Bayesian MCMC samples in Fig  \ref{fig: SAEM_estimates_for_MTT} indicate that the modes of the posterior probabilities obtained using MCMC are around $\theta^{\ast, z}$ as well.

The computational complexity of one MCMC move for updating $z_{1:n}$, for a fixed parameter $\theta$, is dominated by a term which is $\mathcal{O}(\lambda_{x} T^{2} \lambda_{b} \,)$, where $\lambda_{x} = \lambda_{b} / (1- p_{s})$ is the average number of targets per time. 
%Thus one MCMC step for Bayesian estimation has similar complexity to that of MCMC-EM since the cost of update
On the other hand, the cost of the E-step of SMC-EM is dominated by a term which is $\mathcal{O}(T N L \lambda_{y}^{3} )$, where $\lambda_{y} = \lambda_{x}(1 + p_{d}) + \lambda_{f}$ and $L$ is the parameter used in $L$-best assignment. (For a more detailed computational analysis for SMC based EM algorithms see Appendix \ref{sec: Computational complexity of SMC based EM algorithms}.) In realistic scenarios, one expects the SMC E-step, being power three in the number of targets and clutter, to be far more costly then the MCMC E-step, which results in the SMC-EM algorithm being far slower, as in our example. We observed, but not shown in Figure \ref{fig: SAEM_estimates_for_MTT}, that the $\theta$ samples of the MCMC Bayesian estimate reached the true values after approximately $2e4$ iterations, earlier than MCMC-EM's $7.5e4$ iterations. This is because MCMC-EM forgets its past more slowly than MCMC Bayesian due to dependance induced by the stochastic approximation step \eqref{eq: E-step of SAEM}. Although in this case MCMC Bayesian seems preferable, we need to be careful when choosing the prior distribution for $\theta$ especially
when data is scarce as it may unduly influence the results.

\begin{figure*}[t!]
\center{\includegraphics[scale = 0.6]{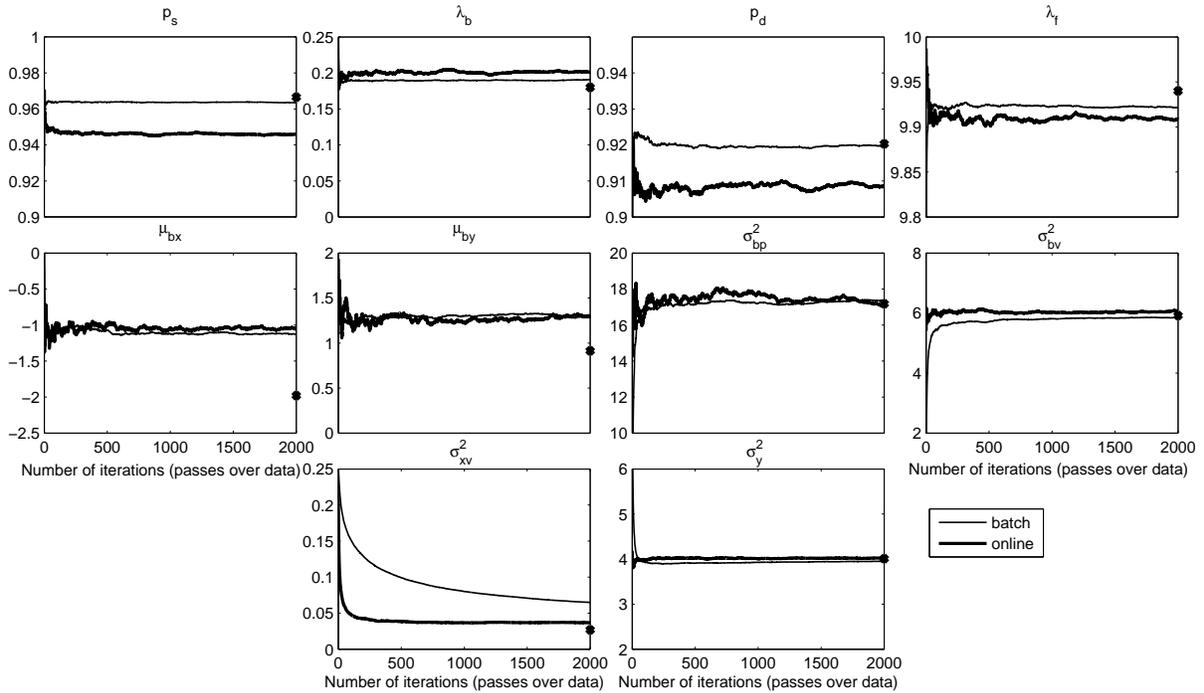}}
\caption{Comparison of online SMC-EM estimates applied to the concatenated data (thicker line) with batch SMC-EM}
\label{fig: Online vs Batch SMC-EM}
\end{figure*}

The reason why SMC-EM is comparatively slow to converge is because of the costly SMC E-step. Often, the parameters can be updated without 
a complete browse through all the data. We may thus speed up convergence by applying SMC online EM (Algorithm \ref{alg: Online EM for the MTT model}) on the following sequence of 
concatenated data 
\[
[ \mathbf{y}_{1:n}, \mathbf{y}_{1:n}, \ldots ],
\] 
Figure \ref{fig: Online vs Batch SMC-EM} shows both our previous SMC-EM estimates (vs number of iterations) in Figure \ref{fig: SAEM_estimates_for_MTT} and the SMC online EM estimates (vs number of passes over the original data $\mathbf{y}_{1:n}$) on the concatenated data; and we note that both algorithms are started with the same initial estimate of $\theta^{\ast}$. Noting that the computational cost of one iteration of the SMC-EM algorithm and the computational cost of one pass of SMC online EM algorithm over the data are roughly the same, we observe that $\sigma_{xv}^2$ and the other parameters converge much quicker in this way. The caveat though is that there is now a bias introduced due to the discontinuity at the concatenation points, e.g.  $\mathbf{y}_{n}$ may correspond to the observations of many surviving targets whereas $\mathbf{y}_{1}$ may be the observations of an initially target free surveillance region. This discontinuity will effect, especially, survival $p_s$, detection $p_d$, and any other parameter depending crucially on a correct $K_{t}^{x}$ estimate over time. However it will have little effect on the parameters $\mu_{bx}, \mu_{by}, \sigma_{bp}^{2}, \sigma_{bv}^{2}, \sigma_{xv}^{2}, \sigma_{y}^{2}$ which govern the dynamics of the HMM associated with a target. In conclusion, one way to estimate $\theta^{\ast}$ in a batch setting using SMC-EM is by (i) first running SMC online EM on $[ \mathbf{y}_{1:n}, \mathbf{y}_{1:n}, \ldots ]$ until convergence to get an estimator $\theta'$ of $\theta^{\ast}$, (ii) and then run the batch SMC-EM initialised at $\theta'$.

\subsubsection{Batch estimation on a larger data set} \label{sec: Batch estimation on a larger data set}
\begin{figure*}[t!]
\centerline{\epsfig{figure=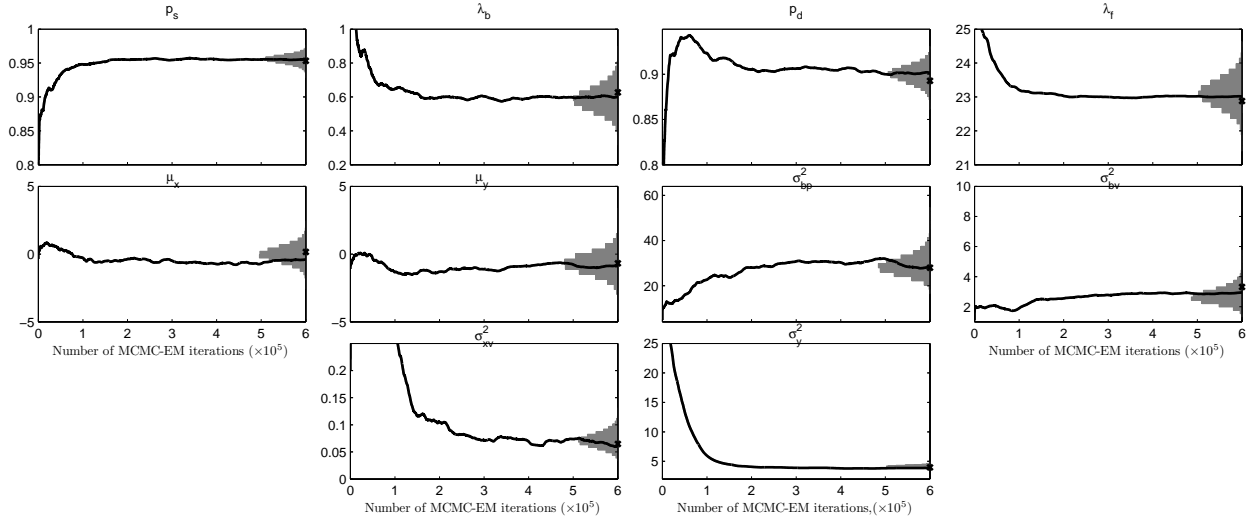, scale = 0.5}}
\caption{Batch estimates obtained from a large data set using the MCMC-EM (bold lines) algorithm for MLE and  MCMC for Bayesian estimates (histograms). $\theta^{\ast, z}$ is shown as a cross. Upper and lower x-axes show the number of EM iterations for MCMC-EM and SMC-EM, respectively.}
\label{fig: SAEM_estimates_for_MTT2}
\end{figure*}

In the second experiment we compare the batch estimation algorithms, MCMC-EM and the Bayesian method, with a larger data set which has more
targets and observations. Recall that the SMC-EM algorithm is based on a SMC algorithm which uses the $L$-best linear assignments and its computational complexity is approximately polynomial of order $3$ in $\lambda_{y} = \lambda_{x} + (1 + p_{d}) \lambda_{f}$.  Therefore, the SMC-EM algorithm would take a long time to execute and is left out of the comparison in this experiment. We created a data set of $n = 150$ time steps by using the parameter
\[
\theta^{\ast} = (0.65, 22.5, 0.90, 0.95, 0, 0, 25, 4, 0, 0.0625, 4).
\]
with window size $\kappa = 150$ for the surveillance region. With this choice, we see approximately $13$ targets per time. Figure \ref{fig: SAEM_estimates_for_MTT2} shows the results obtained from the MCMC-EM and the Bayesian method for estimating $\theta^{\ast}$. When the true association is given, the EM algorithm finds $\theta^{\ast, z}$ for this data set as
\begin{equation*}
\begin{split}
& \theta^{\ast, z} = (0.63, 22.88, 0.90, 0.95, 0.15, -0.68, 27.96, 3.32, \\
& \quad\quad\quad\quad\quad\quad\quad\quad\quad\quad\quad\quad\quad\quad\quad\quad\quad\quad\quad 0, 0.065, 3.98).
\end{split}
\end{equation*}
We can see that both methods work well for this large data set. It is worth mentioning that MCMC Bayesian converged to the stationary distribution after $1e5$ iterations (not shown in the figure), while MCMC-EM converged after $3e5$ iterations.

\subsection{Online EM setting} \label{sec: Online EM setting}
We demonstrate the performance of the SMC online EM in Algorithm \ref{alg: Online EM for the MTT model} in two settings.

\subsubsection{Unknown fixed number of targets} \label{sec: Unknown fixed number of targets}
\begin{figure}[h!]
\centerline{\epsfig{figure=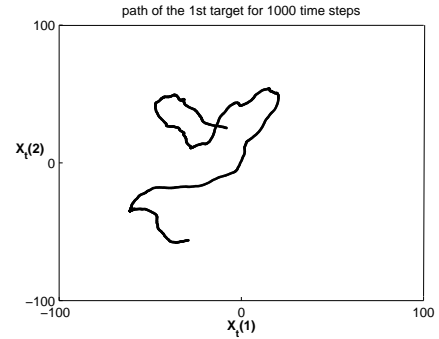, scale = 0.50}}
\caption{The position of target no. 1 evolving in time for the first $1000$ time steps with modified constant velocity model with $F$ in \eqref{modified_F}}
\label{fig: path_of_a_target_in_the_online_setting}
\end{figure}

In the first experiment for online estimation, we create a scenario where there are a constant but unknown number of targets that never die and travel in the surveillance region for a long time. That is, $K_{0}^{x} = K$ (which is unknown and to be estimated), $\lambda_{b} = 0$ and $p_{s} = 1$. We also slightly modify our MTT model so that the target state is a stationary process. The modified model assumes that the state transition matrix $F$ is
\begin{equation}\label{modified_F}
F = \left(\begin{array}{cc} 0.99 I_{2 \times 2} & \Delta I_{2 \times 2} \\ \mathbf{0}_{2 \times 2} & 0.99 I_{2 \times 2} \end{array}\right),
\end{equation}
and $G, W$ and $V$ are the same as the MTT model in Example \ref{ex: The constant velocity model}. The change is to the diagonals of matrix $F$ which should be $I_{2 \times 2}$ for a constant velocity model. However, $0.99 I_{2 \times 2}$ will lead to non-divergent targets, i.e. having a stationary distribution; see Figure \ref{fig: path_of_a_target_in_the_online_setting} for a sample trajectory.
We create data of length $n = 50000$ with $K = 10$ targets which are initiated by using $ \mu_{bx} = 0, \mu_{by} = 0, \sigma_{bx}^{2} = 25, \sigma_{bv}^{2} = 4$. The other parameters to create the data are $p_{d} = 0.9, \lambda_{f} = 10, \sigma_{xv}^{2} = 0.01, \sigma_{y}^{2} = 4$, and the window size $\kappa = 100$.

\begin{figure}[h!]
\centerline{\epsfig{figure=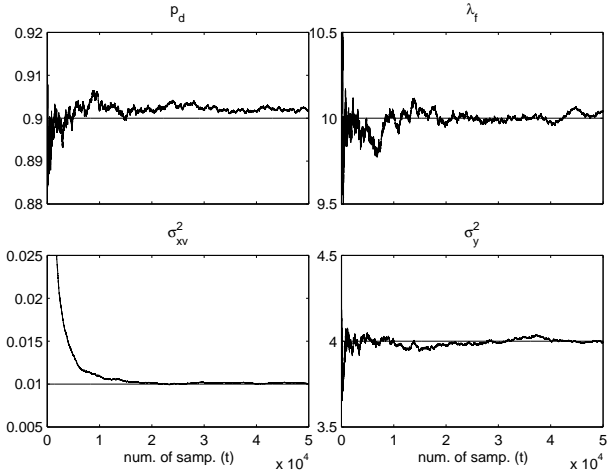, scale = 0.55}}
\caption{Online estimates of SMC-EM algorithm (Algorithm \ref{alg: Online EM for the MTT model}) for fixed number of targets. True values are indicated with a horizontal line. Initial estimates for $p_{d}, \lambda_{f}, \sigma_{xv}^{2}, \sigma_{y}^{2}$ are $0.6, 15, 0.25, 25$; they are not shown in order to zoom in around the converged values.}
\label{fig: online_EM_for_MTT_fixed_num_of_targets}
\end{figure}

Figure \ref{fig: online_EM_for_MTT_fixed_num_of_targets} shows the estimates for parameters $p_{d}, \lambda_{f}, \sigma_{xv}^{2}, \sigma_{y}^{2}$ using the SMC online EM algorithm described in Algorithm \ref{alg: Online EM for the MTT model}, when $K_{t}^{0} = K = 10$ is known. We used $L = 10$ and $N = 100$, and $\gamma_{t} = t^{-0.8}$ is taken for all of the parameters except $\sigma_{xv}^{2}$, where we used $\gamma_{t} = t^{-0.55}$. The burn-in time, until when the M-step is not executed, is $t_{b} = 10$. We can observe the estimates for the parameters quickly settle around the true values.  Note that $\mu_{x},\mu_{y}, \sigma_{bp}^{2}, \sigma_{bv}^{2}$ are not estimated here because they are the parameters of the initial distribution of targets which have no effect on the stationary distribution of a MTT model with fixed number of targets, and thus they are not identifiable by an online EM algorithm \citep{Douc_et_al_2004}.  Note that the online MLE procedure is based on the fact that the parameters of the initial distribution will have a negligible effect on the likelihood of observations $\mathbf{y}_{t}$ for large $t$.
In practice, the parameters of the initial distribution can be estimated by running a batch EM algorithm for the sequence of the first few observations, such as $\mathbf{y}_{1:50}$, and fixing all other parameters to the values obtained by SMC online EM.

\begin{figure}[t!]
\centerline{\hspace{0.5cm} \epsfig{figure=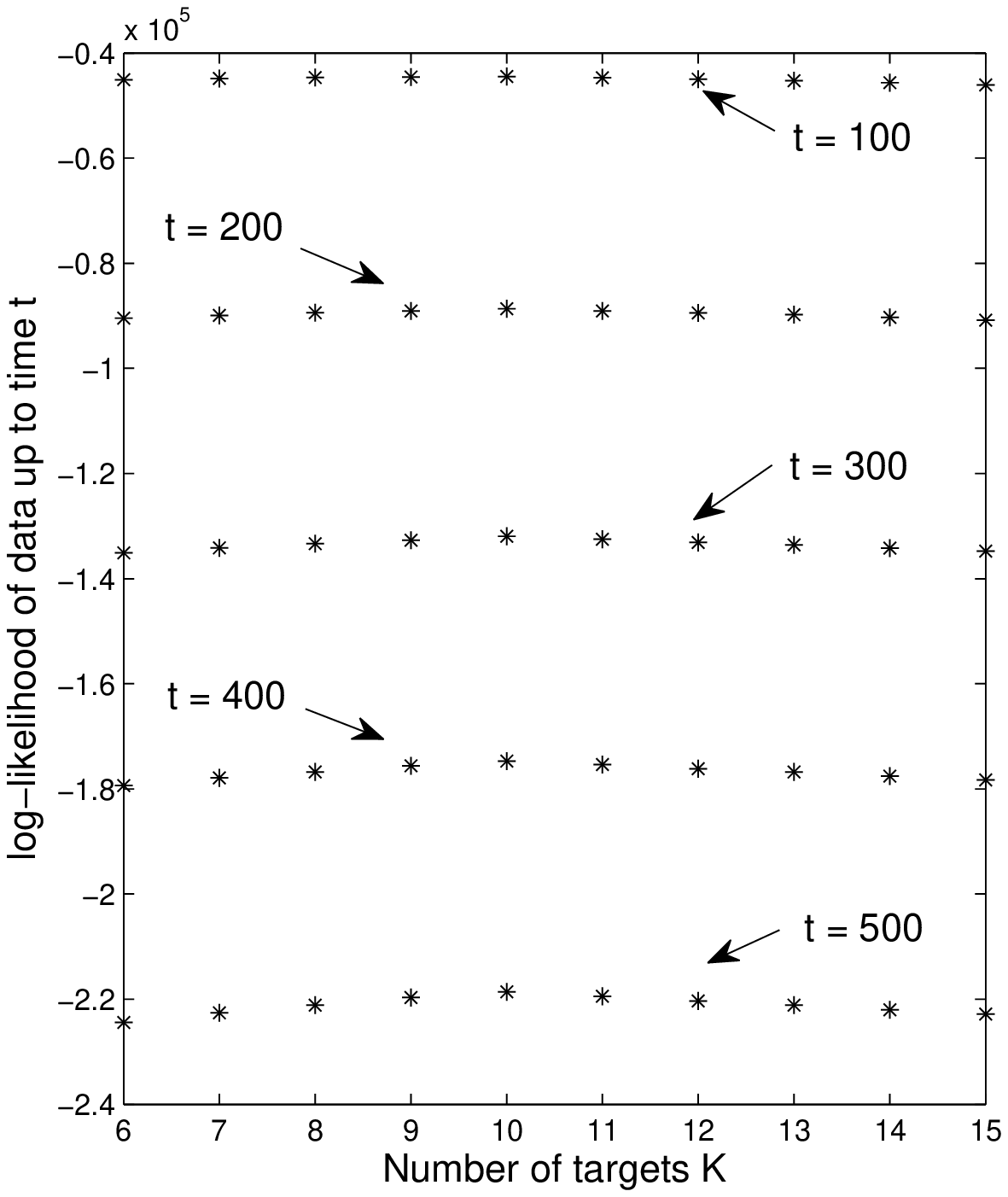, scale = 0.37} \hspace{-0.70cm} \epsfig{figure=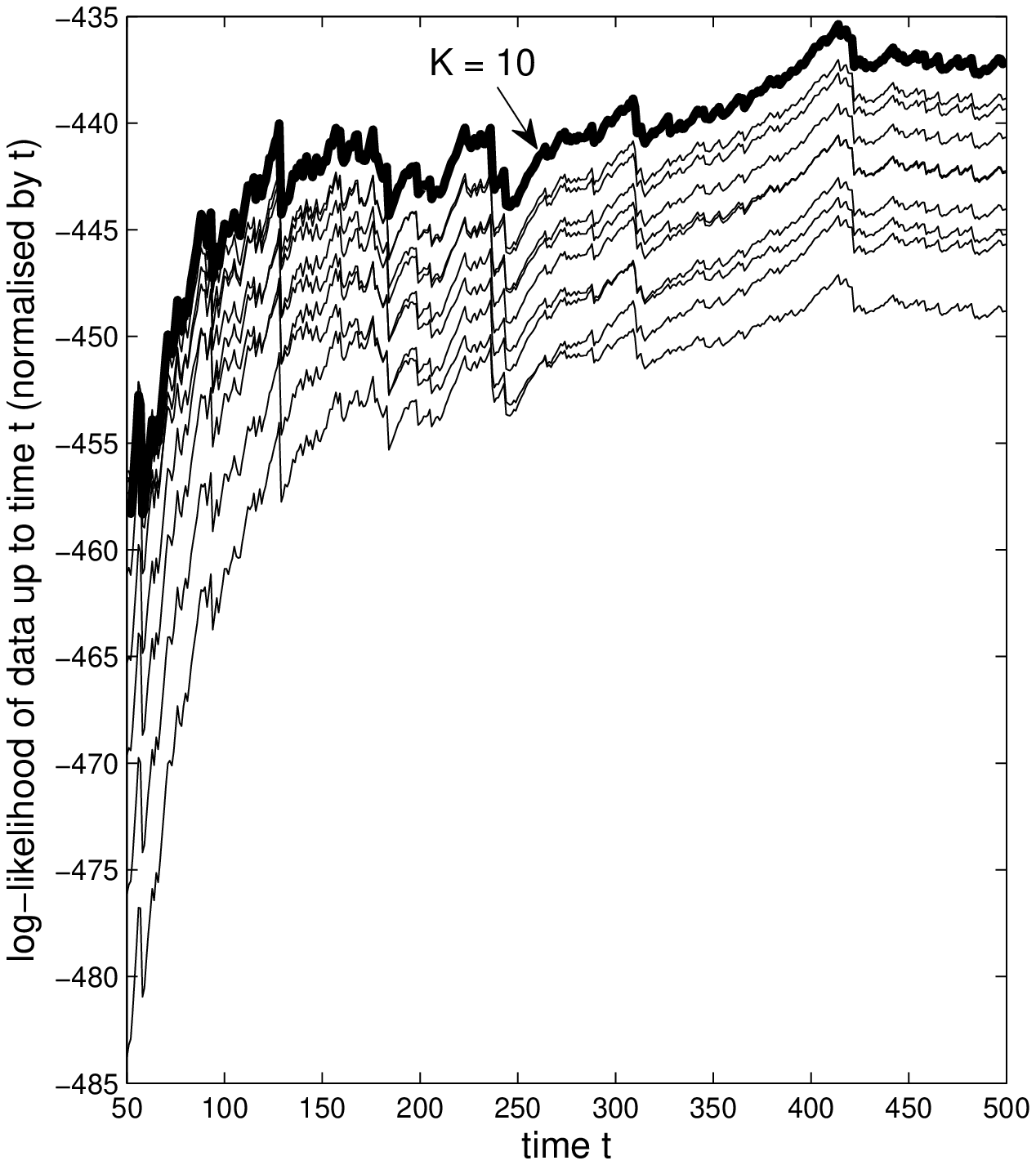, scale = 0.35} }
\caption{Left: estimates of $p_{\theta_{1:t}}(\mathbf{y}_{1:t} | K )$ (normalised by $t$) for values $t = 100 \ldots, t = 500$ and for $K = 6, \ldots, K= 15$. Right: Estimates of $p_{\theta_{1:t}}(\mathbf{y}_{1:t} | K )$ normalised by $t$ for values $K = 6, \ldots, K = 15$, $K = 10$ is stressed with a bold plot.}
\label{fig: likelihoods for selection of number of targets}
\end{figure}

The particle filter in Algorithm \ref{alg: Online EM for the MTT model}, which we used to produce the results in Figure \ref{alg: Online EM for the MTT model}, has all its particles having the same number of targets, which is the true $K$. However, $K$ can be estimated by running several SMC online EM algorithms with different possible $K$'s, and comparing the estimated likelihoods $p_{\theta_{1:t}}(\mathbf{y}_{1:t} | K )$ versus $t$. Figure \ref{fig: likelihoods for selection of number of targets} shows how the estimates of $p_{\theta_{1:t}}(\mathbf{y}_{1:t} | K )$ for values $K = 6, \ldots,15$ compare with time. Both the left and right figures suggest that $p_{\theta_{1:t}}(\mathbf{y}_{1:t} | K )$ favours $K = 10$ starting from $t = 100$ and the decision on the number of targets can be safely made after about $200$ time steps. We have also checked this comparison with different initial values for $\theta$ and found out that the comparison is robust to the initial estimate $\theta_{0}$.

\subsubsection{Unknown time varying number of targets} \label{sec: Unknown and time varying number of targets}
\begin{figure*}[t!]
\centerline{\epsfig{figure=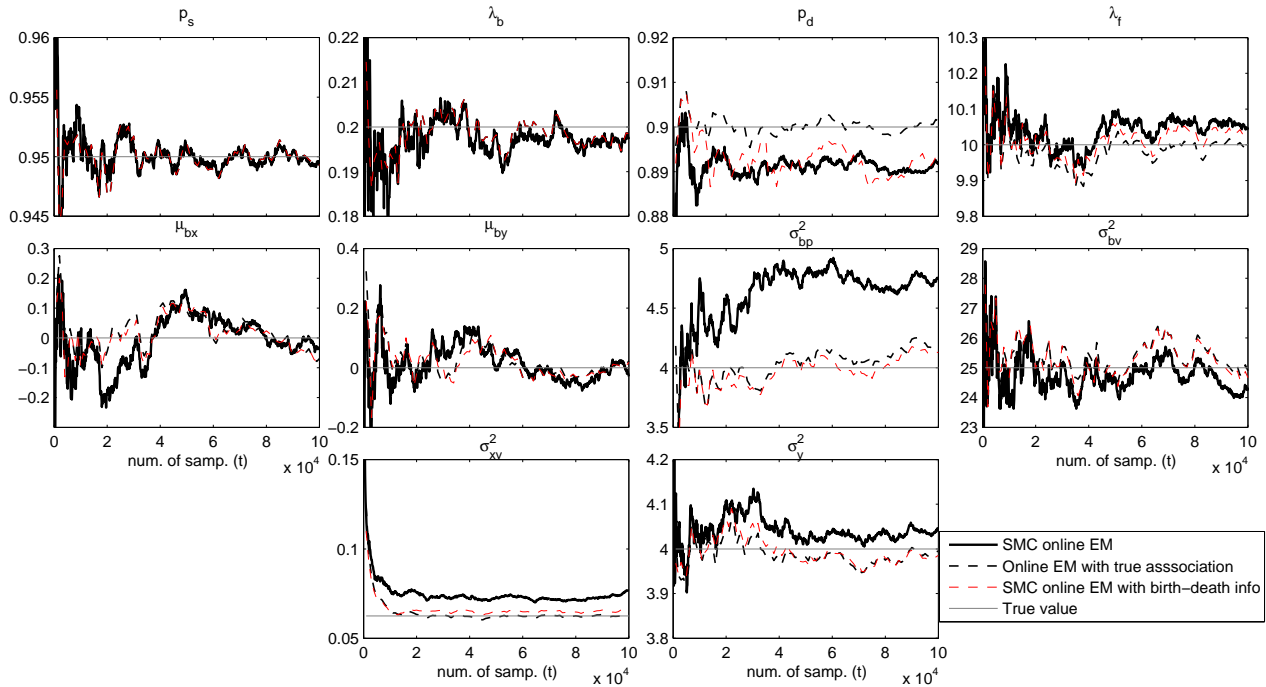, scale = 0.6}}
\caption{Estimates of online SMC-EM algorithm (Algorithm \ref{alg: Online EM for the MTT model}) for a time varying number of targets, compared with online EM estimates when the true data association ($\{ Z_{t} \}_{t \geq 1}$) is known (black dashed lines) and SMC online EM estimates when the birth death information ($\{ K_{t}^{b}, C_{t}^{s} \}_{t \geq 1}$) is known (red dashed lines). For the estimates in case of known true association and in case of known birth-death information, $\theta_{1000, 2000, \ldots, 100000}$ are shown only. True values are indicated with a horizontal line. The initial value $\theta_{0} = (0.8, 0.5, 0.6, 13, -1, -1, 1,  1, 16, 0, 0.25, 25)$ is not shown in order to zoom in around $\theta^{\ast}$}
\label{fig: SMC online EM vs online EM with true data association}
\end{figure*}

In the second experiment with online estimation, we consider the constant velocity model in Example \ref{ex: The constant velocity model} with a time-varying number of targets, i.e. $\lambda_{b} > 0$ and $p_{s} < 1$. We generated a set of data of length $n = 10^5$ using parameters
\[
\theta^{\ast} = \left( 0.2, 10, 0.90, 0.95, 0,  0, 25, 4, 0, 0.0625, 4 \right)
\]
and we estimated all of them (except $\sigma_{xp}^{2} = 0$). Again, we used $L = 10$ and $N = 200$, and $\gamma_{t} = t^{-0.8}$ is taken for all of the parameters except $\sigma_{xv}^{2}$ for which we used $\gamma_{t} = t^{-0.55}$. The online estimates for those parameters are given in Figure \ref{fig: SMC online EM vs online EM with true data association} (solid lines). The initial values are taken to be $\theta_{0} = (0.8, 0.5, 0.6, 13, -1, -1, 1,  1, 16, 0, 0.25, 25)$ which is not shown in the figure in order to zoom in around $\theta^{\ast}$. We observe that the estimates have quickly left their initial values  and settle around $\theta^{\ast}$. Also, the parameter estimates for the initial distribution of newborn targets have the largest oscillations around their true values which is in agreement with the results in the batch setting.

Another important observation from Figure \ref{fig: SMC online EM vs online EM with true data association} is that there is bias in the estimates of some of the parameters, namely $p_{d}, \lambda_{f}, \sigma_{bv}^{2}, \sigma_{xv}^{2}, \sigma_{y}^{2}$. This bias arises from the Monte Carlo approximation. To provide a clearer illustration of this Monte Carlo bias, we compared the SMC online EM estimates with the online EM estimates we would have if we were given the true data association, i.e.\ $\{ Z_{t} \}_{t \geq 1}$. The dashed lines in Figure \ref{fig: SMC online EM vs online EM with true data association} show the results obtained when the true association is known; for illustrative purposes we plot every $1000$'th estimate only, hence the sequence $\theta_{1000, 2000, \ldots, 100000}$. 

The source of the bias in the results is undoubtedly due to the SMC approximation of $p_{\theta}(z_{1:n}|\mathbf{y}_{1:n})$. However, we are able to pin down more precisely which components of $z_{1:n}$ are being poorly tracked. We ran the SMC online EM algorithm for the same data sequence, but this time by feeding the algorithm with the birth-death information, i.e.\ $\{ K_{t}^{b}, C_{t}^{s}  \}_{t \geq 1}$. Figure \ref{fig: SMC online EM vs online EM with true data association} shows that when $\{ K_{t}^{b}, C_{t}^{s}  \}_{t \geq 1}$ is provided to the algorithm, the bias  for some components drops. This indicates that (i) the bias in the MTT parameters is predominantly due to the poor tracking of the birth and death times by our SMC MTT algorithm and (ii) with knowledge of the births and deaths, the unknown assignments of targets to observations seem to be adequately resolved by the $L$-best approach since the bias in the target HMM parameters diminishes. Therefore, the bottle neck of the SMC MTT algorithm is birth/death estimation and, generally speaking, a better SMC scheme for the birth-death tracking may reduce the bias.  Note that when the number of births per time is limited by a finite integer, \textit{all} the variables of $Z_{t}$ i.e.\ $( K_{t}^{b}, K_{t}^{f}, C_{t}^{s}, C_{t}^{d}, A_{t} )$ can be tracked within the $L$-best assignment framework, and we expect in this case the bias to be significantly smaller. However, since in our MTT model the number of births per time is unlimited (being a Poisson random variable), we cannot include birth-death tracking in the $L$-best assignment framework; see the SMC algorithm in Appendix \ref{sec: SMC algorithm for MTT} for details. 

\subsubsection{Tuning the number of particles $N$} \label{sec: Choice of N}
It is expected that a reasonable accuracy of SMC target tracker is necessary for good performance in parameter estimation. Obviously, there is a trade off between accuracy of SMC tracking and computational cost, and this trade off is a function of $N$, the number of particles. This raises the following question: how do we identify if the number of particles is adequate for the SMC online EM algorithm for a real data set given that $\theta^{\ast}$ is unknown? We propose a procedure to address this issue. For the chosen value $N$:
\begin{enumerate}
\item Run SMC online EM on the real data set with $N$ particles to obtain an estimate $\hat{\theta}$ of the unknown $\theta^{\ast}$.
\item Simulate the MTT model with $\hat{\theta}$ for a small number of time steps to obtain a data set for verification.
\item Run the SMC target tracker for the simulated data with $\theta = \hat{\theta}$ known.
\item If the target tracking accuracy is ``bad'', increase $N$ and return to step 1; else stop.
\end{enumerate} 
The tracking accuracy can roughly be measured by comparing $K_{t}^{x}$ with its particle estimate which is suggestive of the birth-death tracking performance, which we have identified to have a significant impact on the bias of the estimates as shown in  Figure \ref{fig: SMC online EM vs online EM with true data association}.

\section{Conclusion and Discussion}\label{sec: Conclusion}
We have presented MLE algorithms for inferring the static parameters in linear Gaussian MTT models. Based on our comparisons of the offline and online EM implementations, our recommendations to the practitioner are: (i) If batch estimation permissible for the application then it should always be preferred. (ii) Moreover, MCMC-EM should be preferred as batch SMC-EM has the disadvantage of slow convergence of some parameters while online SMC-EM applied to concatenated data, although converges quicker then batch MCMC-EM, induces some bias for certain parameters due to the discontinuity caused at the concatenation boundaries. Furthermore, SMC tracker does not scale well with the average number of targets per time and clutter rate; see Sec calculation in \ref{sec: Batch setting}.
 (iii) For very long data sets (i.e. large time) and when there is a computational budget, then online SMC-EM seems the most appropriate since the it is easier to control computational demands by restricting the number of particles. We have seen that in online SMC-EM  there will be biases in some of the parameter estimates if the birth and death times are not tracked accurately. The particle number should be verified for adequacy as recommended in Section \ref{sec: Choice of N}.

 We have not considered other tracking algorithms that work well such as those based on the PHD filter \cite{Vo_and_Ma_2006, Whiteley_et_al_2010} which could be used provided track estimates can be extracted.
The linear Gaussian MTT model can be extended in the following manner while still admitting an EM implementation of MLE. For example, split-merge scenarios for targets can be considered. Moreover, the number of newborn targets per time and false measurements need not be Poisson random variables; for example the model may allow no births or at most one birth at a time determined by a Bernoulli random variable. Furthermore, false measurements need not be uniform, e.g. their distribution may be a Gaussian (or a Gaussian mixture) distribution. Also, we assumed that targets are born close to the centre of the surveillance region; however, different types of initiation for targets may be preferable in some applications.

For non-linear non-Gaussian MTT models, Monte Carlo type batch and online EM algorithms may still be applied by sampling from the hidden states $\mathbf{X}_{t}$'s provided that the sufficient statistics for the EM are available in the required additive form \citep{Del_Moral_et_al_2009}. In those MTT models where sufficient statistics for EM are not available, other methods such as gradient based MLE methods can be useful (e.g.\ \citet{Poyiadjis_et_al_2011}).

\appendix  \label{sec: Appendix}

\subsection{Recursive updates for sufficient statistics in a single GLSSM} \label{sec: Recursive updates for sufficient statistics in a single GLSSM}
Referring to the variables in Section \ref{sec: Online smoothing in a single GLSSM}, the intermediate functions for the sufficient statistics in \eqref{eq: KF sufficient statistics} can be written as
\begin{align}
T_{m, t, ij}(x_{t}, c_{1:t}^{d}) &= x_{t}^{T} \bar{P}_{m, t, ij} x_{t} + \bar{q}_{m, t, ij}^{T} x_{t} + \bar{r}_{m, t, ij} \nonumber
\end{align}
where $i, j = 1, \ldots, d_{x}$ for $m = 1, 3, 4, 5, 7$; $i = 1, \ldots, d_{x},  j = 1, \ldots, d_{y}$ for $m = 2$; and $i = 1, \ldots, d_{x}$, $j = 1$ for $m = 6$.
All $\bar{P}_{m, t, ij}$'s, $\bar{q}_{m, t, ij}$'s and $\bar{r}_{m, t, ij}$'s are $d_{x} \times d_{x}$ matrices, $d_{x} \times 1$ vectors and scalars, respectively.  Forward smoothing is then performed via recursions over these variables. Start at time 1 with the initial conditions  $\bar{P}_{m, 1, ij} = 0_{d_{x} \times d_{x}}$, $\bar{q}_{m, 1, ij} = 0_{d_{x} \times 1}$, and $\bar{r}_{m, 1, ij} = 0$ for all $m$ except $ \bar{P}_{1, 1, ij} = c_{1}^{d} e_{i}e_{j}^{T}$, $\bar{P}_{7, 1, ij} = e_{i}e_{j}^{T}$, $\bar{q}_{2, 1, ij} = c_{1}^{d} y_{1}(j) e_{i}$, and $\bar{q}_{6, 1, i1} = e_{i}$. At time $t+1$, update
\begin{align*} \label{eq: forward smoothing recursion in GLSSM}
\bar{P}_{1, t+1, ij} & = B_{t}^{T} \bar{P}_{1, t, ij} B_{t} + c^{d}_{t+1} e_{i}e_{j}^{T} \\
\bar{q}_{1, t+1, ij} & = B_{t}^{T} \bar{q}_{1, t, ij}  +  B_{t}^{T} \left( \bar{P}_{1, t, ij} +  \bar{P}_{1, t, ij}^{\theta, T} \right) b_{t} \\
\bar{r}_{1, t+1, ij} & = \bar{r}_{1, t, ij} + \text{tr} \left(\bar{P}_{1, t, ij} \Sigma_{t | t+1} \right) + \bar{q}_{1, t, ij}^{T} b_{t} +  b_{t}^{T} \bar{P}_{1, t,ij} b_{t} \\
\bar{P}_{2, t+1, ij} & = 0_{d_{x} \times d_{x}} \\
\bar{q}_{2, t+1, ij} & = B_{t}^{T} \bar{q}_{2, t, ij} + c_{t+1}^{d} y_{t+1}(j) e_{i} \\
\bar{r}_{2, t+1, ij} & = \bar{r}_{2, t, ij} + \bar{q}_{2, t+1, ij}^{T} b_{t} \\
\bar{P}_{3, t+1, ij} & = B_{t}^{T} \left( \bar{P}_{3, t, ij} + e_{i}e_{j}^{T} \right) B_{t} \\
\bar{q}_{3, t+1, ij} & = B_{t}^{T} \bar{q}_{3, t, ij}  +  B_{t}^{T} \left( \bar{P}_{3, t, ij} +  \bar{P}_{3, t, ij}^{T} + e_{i}e_{j}^{T}  + e_{j}e_{i}^{T} \right) b_{t} \\
\bar{r}_{3, t+1, ij} & = \bar{r}_{3, t, ij} + \text{tr} \left( \left( \bar{P}_{3, t, ij} + e_{i}e_{j}^{T} \right) \Sigma_{t | t+1} \right) + \bar{q}_{3, t, ij}^{T} b_{t} \\
& \quad +  b_{t}^{T} \left( \bar{P}_{3, t,ij} + e_{i}e_{j}^{T} \right) b_{t} \\
\bar{P}_{4, t+1, ij} & = B_{t}^{T} \bar{P}_{4, t, ij} B_{t} + e_{i}e_{j}^{T} \\
\bar{q}_{4, t+1, ij} & = B_{t}^{T} \bar{q}_{4, t, ij}  +  B_{t}^{T} \left( \bar{P}_{4, t, ij} +  \bar{P}_{4, t, ij}^{T} \right) b_{t} \\
\bar{r}_{4, t+1, ij} & = \bar{r}_{4, t, ij} + \text{tr} \left(\bar{P}_{4, t, ij} \Sigma_{t | t+1} \right) + \bar{q}_{4, t, ij}^{T} b_{t} +  b_{t}^{T} \bar{P}_{4, t,ij} b_{t} \\
\bar{P}_{5, t+1, ij} & = B_{t}^{T} \bar{P}_{5, t, ij} B_{t} + e_{i}e_{j}^{T} B_{t} \\
\bar{q}_{5, t+1, ij} & = B_{t}^{T} \bar{q}_{5, t, ij}  +  B_{t}^{T} \left( \bar{P}_{5, t, ij} +  \bar{P}_{5, t, ij}^{T} \right) b_{t} + e_{j} b_{k}^{T} e_{i} \\
\bar{r}_{5, t+1, ij} & = \bar{r}_{5, t, ij} + \text{tr} \left(\bar{P}_{5, t, ij} \Sigma_{t | t+1} \right) + \bar{q}_{5, t, ij}^{T} b_{t} +  b_{t}^{T} \bar{P}_{5, t,ij} b_{t} \\
\bar{P}_{6, t+1, i1} & = 0_{d_{x} \times d_{x}} \\
\bar{q}_{6, t+1, i1} & = B_{t}^{T} \bar{q}_{6, t, i1} \\
\bar{r}_{6, t+1, i1} & = \bar{r}_{6, t, i1} + \bar{q}_{6, t+1, i1}^{T} b_{t} \\
\bar{P}_{7, t+1, ij} & = B_{t}^{T} \left( \bar{P}_{7, t, ij} \right) B_{t} \\
\bar{q}_{7, t+1, ij} & = B_{t}^{T} \bar{q}_{7, t, ij}  +  B_{t}^{T} \left( \bar{P}_{7, t, ij} +  \bar{P}_{7, t, ij}^{T} \right) b_{t} \\
\bar{r}_{7, t+1, ij} & = \bar{r}_{7, t, ij} + \text{tr} \left( \bar{P}_{7, t, ij}  \Sigma_{t | t+1} \right) + \bar{q}_{7, t, ij}^{T} b_{t} +  b_{t}^{T}  \bar{P}_{7, t,ij} b_{t}
\end{align*}

For the online EM algorithm, we simply modify the update rules by multiplying the terms on the right hand side containing $e_{t}$ or $I_{d_{x} \times d_{x}}$ by $\gamma_{t+1}$ and multiplying the rest of the terms by $(1 - \gamma_{t+1})$.

\subsection{SMC algorithm for MTT} \label{sec: SMC algorithm for MTT}
An SMC algorithm is mainly characterised by its proposal distribution. Hence, in this section we present the proposal distribution $q_{\theta}(z_{t} | z_{1:t-1}, \mathbf{y}_{1:t} )$, where we exclude the superscripts for particle numbers from the notation for simplicity. Assume that $z_{1:t-1}$ is the ancestor of the particle of interest with weight $w_{t-1}$. We sample $z_{t} = (k_{t}^{b}, c_{t}^{s}, c_{t}^{d}, k_{t}^{f}, a_{t} )$ and calculate its weight by performing the following steps:
\begin{itemize}
\item \textit{Birth-death move:} Sample $k_{t}^{b} \sim \mathcal{PO}(\cdot; \lambda_{b})$ and $c_{t}^{s}(j) \sim \mathcal{BE}(\cdot; p_{s})$ for $j = 1, \ldots, k_{t-1}^{x}$. Set $k_{t}^{s} = \sum_{j = 1}^{k_{t-1}^{x}} c_{t}^{s}$ and construct the $k_{t}^{s} \times 1$ vector $i_{t}^{s}$ from $c_{t}^{s}$. Set $k_{t}^{x} = k_{t}^{s} + k_{t}^{b}$ and calculate the prediction moments for the state. For $j = 1, \ldots, k_{t}^{x}$,
\begin{itemize}
\item if $j \leq k_{t}^{s}$, set $\mu_{t|t-1, j} = F \mu_{t-1 | t-1, i_{t}^{s}(j)}$ and  $\Sigma_{t | t-1, j} = F \Sigma_{t-1 | t-1, i_{t}^{s}(j)} F^{T} + W$.
\item if $j > k_{t}^{s}$, set $\mu_{t|t-1, j} = \mu_{b}$ and $\Sigma_{t | t-1, j}  = \Sigma_{b}$.
\end{itemize}
Also, calculate the moments of the conditional observation likelihood: For $j = 1, \ldots, k_{t}^{x}$,
$\mu_{t, j}^{y} = G \mu_{t|t-1, j}$ and $\Sigma_{t, j}^{y} = G \Sigma_{t | t-1, j} G^{T} + V$.

\item \textit{Detection and association}
Define the $k^{x}_{t} \times ( k^{y}_{t} + k^{x}_{t} )$ matrix $D_{t}$ as
\begin{equation*}
D_{t}(i, j) =
\begin{cases}
     \log (p_{d} \mathcal{N}(y_{t, i}; \mu_{t, j}^{y}, \Sigma_{t, j}^{y}) ) & \text{ if } j \leq k^{y}_{t}, \\
    \log \frac{(1- p_{d}) \lambda_{f}}{| \mathcal{Y} |}  & \text{ if } i = j  - k^{y}_{t}, \\
    - \infty & \text{ otherwise. }
\end{cases}
\end{equation*}
and an assignment is a \emph{one-to-one} mapping
$ \alpha_{t} : \{ 1, \ldots, k^{x}_{t} \} \rightarrow \{ 1, \ldots, k^{y}_{t} + k^{x}_{t} \}$. The cost of the assignment, up to an identical additive constant for each $\alpha_{t}$ is
\[
d(D_{t}, \alpha_{t}) = \sum_{j = 1}^{k_{t}^{d}} D_{t}(j, \alpha_{t}(j)).
\]
Find the set $\mathcal{A}_{L} = \{ \alpha_{t, 1}, \ldots, \alpha_{t, L} \}$ of $L$ assignments producing the highest assignment scores. The set $\mathcal{A}_{L}$ can be found using the Murty's assignment ranking algorithm \citep{Murty_1968}. Finally, sample $\alpha_{t} = \alpha_{t, j}$ with probability
\[
\kappa(\alpha_{t, j}) = \frac{\exp[d(D_{t}, \alpha_{t, j})]}{\sum_{j^{\prime} = 1}^{L} \exp[d(D_{t}, \alpha_{t, j^{\prime}})]}, \quad j = 1, \ldots, L
\]
Given $\alpha_{t}$, one can infer $c^{d}_{t}$ (hence $i^{d}_{t}$), $k^{d}_{t}$, $k^{f}_{t}$ and the association $a_{t}$ as follows:
\begin{equation*}
c^{d}_{t}(k) =
\begin{cases}
     1 & \text{ if } \alpha_{t}(k) \leq k^{y}_{t}, \\
     0 & \text{ if } \alpha_{t}(k) > k^{y}_{t}.
\end{cases}
\end{equation*}
Then $k^{d}_{t} = \sum_{j = 1}^{k^{x}_{t}} c^{d}_{t}(k) $, $k^{f}_{t} = k^{y}_{t} - k^{d}_{t}$, $i^{d}_{t}$ is constructed from $c^{d}_{t}$, and finally
\[
a_{t}(k) = \alpha_{t}(i^{d}_{t}(k)), \quad k = 1, \ldots, k^{d}_{t}.
\]
\item \textit{Reweighting:} After we sample $z_{t} = \left(k_{t}^{b}, c_{t}^{s}, c_{t}^{d}, k_{t}^{f}, a_{t} \right)$ from $q_{\theta}(z_{t} | z_{1:t-1}, \mathbf{y}_{t})$, we calculate the weight of the particle as in \eqref{eq: particle weight}, which becomes for this sampling scheme as
\[
w_{t} \propto w_{t-1} \lambda_{f}^{- k^{x}_{t}} \sum_{j = 1}^{L} \exp[d(D_{t}, \alpha_{t, j})].
\]
\end{itemize}

\subsection{Computational complexity of SMC based EM algorithms} \label{sec: Computational complexity of SMC based EM algorithms}

\subsubsection{Computational complexity of SMC filtering} \label{sec: Computational complexity of SMC filtering}
For simplicity, assume the true parameter value is $\theta$. The computational cost of SMC filtering with $\theta$ and $N$ particles, at time $t$, is
\begin{align*}
& C_{\text{SMC}}(\theta, t, N) = \underbrace{c_{1} N}_{\text{resampling}} + \sum_{i = 1}^{N} \Bigg[ \underbrace{\left( c_{2} K_{t-1}^{x (i)} + c_{3} \right)}_{\text{birth-death sampling}} \nonumber \\
& \quad\quad\quad\quad\quad + \underbrace{d_{x}^{3} \left( c_{4} K_{t}^{x}  + c_{5} K_{t}^{x} K_{t}^{y} \right) }_{\text{moments and assignments}} + \underbrace{c_{6} L \left( K_{t}^{x (i)} + K_{t}^{y} \right)^{3} }_{\text{Murty (worst case)}} \Bigg] 
\end{align*}
where $c_{1}$ to $c_{6}$ are constants and $c_{3}$ is for sampling from the Poisson distribution. If we assume that SMC tracks the number of births and deaths well on average then we can simplify the term above
\begin{align*}
C_{\text{SMC}}(\theta, t, N) & \approx N \Big[   c_{1, 3} + c_{2} K_{t-1}^{x} \nonumber \\
& + d_{x}^{3} \left( c_{4} K_{t}^{x}  + c_{5} K_{t}^{x} K_{t}^{y} \right) + c_{6} L \left( K_{t}^{x} + K_{t}^{y} \right)^{3} \Big] 
\end{align*} 
where $c_{1, 3} = c_{1} + c_{3}$. The process $\{ K_{t}^{x} \}_{t \geq 1}$ is Markov and its stationary distribution is $\mathcal{P}(\lambda_{x})$ where $\lambda_{x} = \frac{\lambda_{b}}{1 - p_{s}}$. Also $K_{t}^{y} = K_{t}^{d} + K_{t}^{f}$ and for simplicity we write $K_{t}^{d} \approx p_{d} K_{t}^{x}$. Therefore the stationary distribution for $\{ K_{t}^{x} +  K_{t}^{y}  \}_{t \geq 1}$ is approximately that of $\{ (1 + p_{d}) K_{t}^{x} + K_{t}^{f} \}_{t \geq 1}$ which is $\mathcal{P}(\lambda_{y})$ where $\lambda_{y} = \lambda_{x} (1 + p_{d}) + \lambda_{f}$. Therefore, assuming stationarity at time $t$ and substituting $\mathbb{E}_{\mathcal{P}(\lambda)}(X^{3}) = \lambda^{3} + 3 \lambda^{2} + \lambda$, the expected cost will be
\begin{align*}
\mathbb{E}_{\theta} \left[ C_{\text{SMC}}(\theta, t, N) \right]  &\approx N \Big[ c_{1, 3} + \left( c_{2} + d_{x}^{3}  \left[ c_{4} + c_{5} \left( p_{d} + \lambda_{f} \right)  \right]  \right) \lambda_{x}  \nonumber \\
& \quad\quad\quad + c_{5} p_{d} \lambda_{x}^{2}  + c_{6} L \left( \lambda_{y}^{3} + 3 \lambda_{y}^{2} + \lambda_{y} \right) \Big].
\end{align*}

\subsubsection{SMC-EM for the batch setting} \label{sec: SMC-EM for the batch setting}
The SMC-EM algorithm for the batch setting first runs the SMC filter, stores all its path trajectories i.e.\ $\{ Z_{1:n}^{(i)} \}_{1 \leq i \leq N}$ and then calculates the estimates of required sufficient statistics for each $Z_{1:n}^{(i)}$ by using a forward filtering backward smoothing (FFBS) technique, which is bit quicker then forward smoothing. Therefore, the overall expected cost of batch SMC-EM applied to data of size $n$ is 
\[
C_{\text{SMC-EM}} = C_{\text{FFBS}}(\theta, n, N) + \sum_{t = 1}^{n} C_{\text{SMC}}(\theta, t, N) + c_{7}
\]
where $c_{7}$ is the cost of the M-step, i.e.\ $\Lambda$. Let us denote the total number of targets up to time $n$ is $M$ and let $L_{1}, \ldots, L_{M}$ be their life lengths. The computational cost of FFBS to calculate the smoothed estimates of sufficient statistics for a target of life length $L$ is $\mathcal{O}(d_{x}^{3} L)$. Therefore,
\[
C_{\text{FFBS}}(\theta, n, N) = \sum_{i = 1}^{N} \sum_{m = 1}^{M^{(i)}} c_{8} d_{x}^{3} L_{m}^{(i)}.
\]
Assume the particle filter tracks well and $M^{(i)}$ and $L_{m}^{(i)}$, $m = 1, \ldots, M^{(i)}$ for particles $i = 1, \ldots, N$ are close enough to $L_{m}$, and $M$, the true values, for $m = 1, \ldots, M$. Then, we have 
\[
C_{\text{FFBS}}(\theta, n, N) \approx \sum_{i = 1}^{N} \sum_{m = 1}^{M} c_{8} d_{x}^{3} L_{m}.
\]
The expected values of $L_{m}$ and $M$ are $1/(1 - p_{s})$, $n \lambda_{b}$, respectively. Also assume stationarity at all times so that the expectations of the terms $C_{\text{SMC}}(\theta, t, N)$ are the same and we have
\[
\mathbb{E}_{\theta} \left[ C_{\text{FFBS}}(\theta, n, N) \right] \approx c_{8} N n d_{x}^{3} \lambda_{b}(1 - p_{s})^{-1}.
\]
As a result, given a data set of $n$ time points, the overall expected cost of SMC-EM for the batch setting per iteration is 
\begin{align*}
& \mathbb{E}_{\theta} \left[ C_{\text{SMC-EM}} \right]  \approx \mathbb{E}_{\theta} \left[ C_{\text{FFBS}}(\theta, n, N) \right]  +  n \mathbb{E}_{\theta} \left[ C_{\text{SMC}}(\theta, t, N) \right] + c_{7}. \nonumber  % \\
%& \quad\quad\quad = n N \Big\{ c_{1, 3} + \left( c_{2} + d_{x}^{3} \left[ c_{4} + c_{8} + c_{5} (p_{d} + \lambda_{f}) \right] \right) \lambda_{x} \nonumber \\
% & \quad\quad\quad\quad\quad\quad\quad + c_{5} d_{x}^{3} p_{d} \lambda_{x}^{2} + c_{6} L \left( \lambda_{y}^{3} + 3 \lambda_{y}^{2} + \lambda_{y} \right) \Big\} + c_{8} .
\end{align*}

\subsubsection{SMC online EM} \label{sec: SMC online EM}
The overall cost of an SMC online EM for a data set of $n$ time points is 
\[
C_{\text{SMConEM}} \approx  \sum_{t = 1}^{n} \left[ C_{\text{FSR}}(\theta, t, N) +  C_{\text{SMC}}(\theta, t, N) + c_{7} \right].
\]
The forward smoothing recursion and maximisation used in the SMC online EM requires 
\[
C_{\text{FSR}}(\theta, t, N) = \sum_{i = 1}^{N} c_{9} K_{t}^{x(i)} d_{x}^{5}
\]
calculations at time $t$ for a constant $c_{9}$, whose expectation is 
\[
\mathbb{E}_{\theta} \left[ C_{\text{FSR}}(\theta, t, N) \right]  = c_{9} N \lambda_{b}(1 - p_{s})^{-1} d_{x}^{5}.
\] 
at stationarity. The overall expected cost of an SMC online EM for a data of $n$ time steps, assuming stationarity, is
\begin{align*}
& \mathbb{E}_{\theta} \left[ C_{\text{SMConEM}}(\theta, n, N) \right] \nonumber \\
& \quad\quad\quad\quad \approx n \left( \mathbb{E}_{\theta} \left[ C_{\text{FSR}}(\theta, t, N) \right] + \mathbb{E}_{\theta} \left[ C_{\text{SMC}}(\theta, t, N) \right] + c_{7} \right).  \nonumber % \\
% & = n \left[ N \left\{ c_{1, 3} + \left( c_{2} + d_{x}^{3} \left[ c_{4} + c_{5} (p_{d} + \lambda_{f}) \right] + c_{9} d_{x}^{5} \right) \lambda_{x} + c_{5} d_{x}^{3} p_{d} \lambda_{x}^{2} + c_{6} L \left( \lambda_{y}^{3} + 3 \lambda_{y}^{2} + \lambda_{y} \right)  \right\} + c_{7} \right].
\end{align*}

\bibliographystyle{plainnat}
{\footnotesize
\bibliography{myrefs_paper}}

\begin{thebibliography}{34}
\providecommand{\natexlab}[1]{#1}
\providecommand{\url}[1]{\texttt{#1}}
\expandafter\ifx\csname urlstyle\endcsname\relax
  \providecommand{\doi}[1]{doi: #1}\else
  \providecommand{\doi}{doi: \begingroup \urlstyle{rm}\Url}\fi

\bibitem[Bar-Shalom and Fortmann(1988)]{Bar-Shalom_and_Fortmann_1988}
Yaakov Bar-Shalom and Thomas~E. Fortmann.
\newblock \emph{Tracking and Data Association}.
\newblock Academic Press, Boston:, 1988.
\newblock ISBN 0120797607.

\bibitem[Bar-Shalom and Li(1995)]{Bar-Shalom_and_Li_1995}
Yaakov Bar-Shalom and X.R. Li.
\newblock \emph{Multitarget-Multisensor Tracking: Principles and Techniques}.
\newblock {Y}{B}{S} {P}ublishig, 1995.
\newblock ISBN 0120797607.

\bibitem[Capp{\'e}(2009)]{Cappe_2009a}
O~Capp{\'e}.
\newblock Online sequential {M}onte {C}arlo {E}{M} algorithm.
\newblock In \emph{Proc. IEEE Workshop on Statistical Signal Processing}, 2009.

\bibitem[Capp{\'e}(2011)]{Cappe_2011}
O~Capp{\'e}.
\newblock Online {E}{M} algorithm for hidden {M}arkov models.
\newblock \emph{Journal of Computational and Graphical Statistics}, 20\penalty0
  (3):\penalty0 728--749, 2011.

\bibitem[Celeux and Diebolt(1985)]{Celeux_and_Diebolt_1985}
G.~Celeux and J.~Diebolt.
\newblock The {S}{E}{M} algorithm: A probabilistic teacher algorithm derived
  from the {E}{M} algorithm for the mixture problem.
\newblock \emph{Computational Statistics Quarterly}, 2:\penalty0 73--82, 1985.

\bibitem[Cox and Miller(1995)]{Cox_and_Miller_1995}
Ingemar~J. Cox and Matt~L. Miller.
\newblock On finding ranked assignments with application to multi-target
  tracking and motion correspondence.
\newblock \emph{IEEE Trans. on Aerospace and Electronic Systems}, 32:\penalty0
  48--9, 1995.

\bibitem[Danchick and Newnam(2006)]{Danchick_and_Newnam_2006}
R.~Danchick and G.~E. Newnam.
\newblock Reformulating {R}eid's {M}{H}{T} method with generalised {M}urty
  {K}-best ranked linear assignment algorithm.
\newblock \emph{IEE Proceedings - Radar, Sonar and Navigation}, 153\penalty0
  (1):\penalty0 13--22, 2006.
\newblock \doi{10.1049/ip-rsn:20050041}.
\newblock URL \url{http://link.aip.org/link/?IRS/153/13/1}.

\bibitem[Del~Moral et~al.(2009)Del~Moral, Doucet, and
  Singh]{Del_Moral_et_al_2009}
P.~Del~Moral, A.~Doucet, and S.S Singh.
\newblock Forward smoothing using sequential {M}onte {C}arlo.
\newblock Technical Report 638, Cambridge University, Engineering Department,
  2009.

\bibitem[Delyon et~al.(1999)Delyon, Lavielle, and Moulines]{Delyon_et_al_1999}
Bernard Delyon, Marc Lavielle, and Eric Moulines.
\newblock Convergence of a stochastic approximation version of the {E}{M}
  algorithm.
\newblock \emph{The Annals of Statistics}, 27\penalty0 (1):\penalty0 pp.
  94--128, 1999.
\newblock ISSN 00905364.
\newblock URL \url{http://www.jstor.org/stable/120120}.

\bibitem[Douc et~al.(2004)Douc, \'{E}ric Moulines, and
  Ryd\'{e}n]{Douc_et_al_2004}
Randal Douc, \'{E}ric Moulines, and Tobias Ryd\'{e}n.
\newblock Asymptotic properties of the maximum likelihood estimator in
  autoregressive models with {M}arkov regime.
\newblock \emph{Ann. Statist.}, 32\penalty0 (5):\penalty0 2254--2304, 2004.

\bibitem[Doucet et~al.(2000)Doucet, Godsill, and Andrieu]{Doucet_et_al_2000}
A.~Doucet, S.J. Godsill, and C.~Andrieu.
\newblock On sequential {M}onte {C}arlo sampling methods for {B}ayesian
  filtering.
\newblock \emph{Statistics and Computing}, 10:\penalty0 197--208, 2000.

\bibitem[Elliott and Krishnamurthy(1999)]{Elliott_and_Krishnamurthy_1999}
R.J. Elliott and V.~Krishnamurthy.
\newblock New finite-dimensional filters for parameter estimation of
  discrete-time linear {G}aussian models.
\newblock \emph{Automatic Control, IEEE Transactions on}, 44\penalty0
  (5):\penalty0 938 --951, may. 1999.
\newblock ISSN 0018-9286.
\newblock \doi{10.1109/9.763210}.

\bibitem[Elliott et~al.(2002)Elliott, Ford, and Moore]{Elliott_et_al_2002}
Robert~J. Elliott, Jason~J. Ford, and John~B. Moore.
\newblock On-line almost-sure parameter estimation for partially observed
  discrete-time linear systems with known noise characteristics.
\newblock \emph{International Journal of Adaptive Control and Signal
  Processing}, 16:\penalty0 435--453, 2002.
\newblock \doi{10.1002/acs.703}.

\bibitem[Hue et~al.(2002)Hue, Le~Cadre, and Perez]{Hue_et_al_2002}
C.~Hue, J.-P. Le~Cadre, and P.~Perez.
\newblock Sequential {M}onte {C}arlo methods for multiple target tracking and
  data fusion.
\newblock \emph{Signal Processing, IEEE Transactions on}, 50\penalty0
  (2):\penalty0 309--325, feb 2002.
\newblock ISSN 1053-587X.
\newblock \doi{10.1109/78.978386}.

\bibitem[Mahler(2003)]{Mahler_2003}
R.P.S. Mahler.
\newblock Multitarget {B}ayes filtering via first-order multitarget moments.
\newblock \emph{Aerospace and Electronic Systems, IEEE Transactions on},
  39\penalty0 (4):\penalty0 1152 -- 1178, oct. 2003.
\newblock ISSN 0018-9251.
\newblock \doi{10.1109/TAES.2003.1261119}.

\bibitem[Mahler et~al.(2011)Mahler, Vo, and Vo]{Mahler_et_al_2011}
R.P.S. Mahler, B.T. Vo, and B.N. Vo.
\newblock {C}{P}{H}{D} filtering with unknown clutter rate and detection
  profile.
\newblock \emph{Signal Processing, IEEE Transactions on}, 59\penalty0
  (8):\penalty0 3497--3513, 2011.

\bibitem[Mongillo and Deneve(2008)]{Mongillo_and_Deneve_2008}
G.~Mongillo and S.~Deneve.
\newblock Online learning with hidden {M}arkov models.
\newblock \emph{Neural Computation}, 20\penalty0 (7):\penalty0 1706--1716,
  2008.

\bibitem[Murty(1968)]{Murty_1968}
Katta~G. Murty.
\newblock An algorithm for ranking all the assignments in order of increasing
  cost.
\newblock \emph{Operations Research}, 16\penalty0 (3):\penalty0 682--687, 1968.
\newblock URL \url{http://www.jstor.org/stable/168595}.

\bibitem[Ng et~al.(2005)Ng, Li, Godsill, and Vermaak]{Ng_et_al_2005}
W.~Ng, J.~Li, S.~Godsill, and J.~Vermaak.
\newblock A hybrid approach for online joint detection and tracking for
  multiple targets.
\newblock In \emph{Aerospace Conference, 2005 IEEE}, pages 2126 --2141, march
  2005.
\newblock \doi{10.1109/AERO.2005.1559504}.

\bibitem[Oh et~al.(2009)Oh, Russell, and Sastry]{Oh_et_al_2009}
Songhwai Oh, S.~Russell, and S.~Sastry.
\newblock Markov chain {M}onte {C}arlo data association for multi-target
  tracking.
\newblock \emph{Automatic Control, IEEE Transactions on}, 54\penalty0
  (3):\penalty0 481 --497, march 2009.
\newblock ISSN 0018-9286.
\newblock \doi{10.1109/TAC.2009.2012975}.

\bibitem[Poyiadjis et~al.(2011)Poyiadjis, Doucet, and
  Singh]{Poyiadjis_et_al_2011}
George Poyiadjis, Arnaud Doucet, and Sumeetpal~S. Singh.
\newblock Particle approximations of the score and observed information matrix
  in state space models with application to parameter estimation.
\newblock \emph{Biometrika}, 2011.
\newblock \doi{10.1093/biomet/asq062}.

\bibitem[Reid(1979)]{Reid_1979}
Donald~B. Reid.
\newblock An algorithm for tracking multiple targets.
\newblock \emph{IEEE Transactions on Automatic Control}, 24:\penalty0 843--854,
  1979.

\bibitem[Singh et~al.(2011)Singh, Whiteley, and Godsill]{Singh_et_al_2011}
S.~Singh, N.~Whiteley, and S.~Godsill.
\newblock An approximate likelihood method for estimating the static parameters
  in multi-target tracking models.
\newblock In D.~Barber, T.~Cemgil, and S.~Chiappa, editors, \emph{Bayesian Time
  Series Models}, chapter~11, pages 225--244. Cambridge University Press, 2011.

\bibitem[Singh et~al.(2009)Singh, Vo, Baddeley, and Zuyev]{Singh_et_al_2009}
Sumeetpal~S. Singh, Ba-Ngu Vo, Adrian Baddeley, and Sergei Zuyev.
\newblock Filters for spatial point processes.
\newblock \emph{SIAM J. Control Optim.}, 48\penalty0 (4):\penalty0 2275--2295,
  June 2009.
\newblock ISSN 0363-0129.
\newblock \doi{10.1137/070710457}.
\newblock URL \url{http://dx.doi.org/10.1137/070710457}.

\bibitem[Storlie et~al.(2009)Storlie, Lee, Hannig, and
  Nychka]{Storlie_et_al_2009}
C.B. Storlie, T.C. Lee, J.~Hannig, and D.W. Nychka.
\newblock Tracking of multiple merging and splitting targets: {A} statistical
  perspective.
\newblock \emph{Statistica Sinica}, 19:\penalty0 1--52, 2009.

\bibitem[Streit and Luginbuhi(1995)]{Streit_and_Luginbuhi_1995}
R.~Streit and T.~Luginbuhi.
\newblock Probabilistic multi-hypothesis tracking.
\newblock Technical Report 10,428, Naval Undersea Warfare Center Division,
  Newport, Rhode Island, February 1995.

\bibitem[Vermaak et~al.(2005)Vermaak, Godsill, and Perez]{Vermaak_et_al_2005}
J.~Vermaak, S.J. Godsill, and P.~Perez.
\newblock Monte {C}arlo filtering for multi target tracking and data
  association.
\newblock \emph{Aerospace and Electronic Systems, IEEE Transactions on},
  41\penalty0 (1):\penalty0 309 -- 332, jan. 2005.
\newblock ISSN 0018-9251.
\newblock \doi{10.1109/TAES.2005.1413764}.

\bibitem[Vo et~al.(2005)Vo, Singh, and Doucet]{Vo_et_al_2005}
B.-N. Vo, S.~Singh, and A.~Doucet.
\newblock Sequential {M}onte {C}arlo methods for multitarget filtering with
  random finite sets.
\newblock \emph{Aerospace and Electronic Systems, IEEE Transactions on},
  41\penalty0 (4):\penalty0 1224 -- 1245, oct. 2005.
\newblock ISSN 0018-9251.
\newblock \doi{10.1109/TAES.2005.1561884}.

\bibitem[Vo et~al.(2003)Vo, Singh, and Doucet]{Vo_et_al_2003}
Ba-Ngu Vo, S.~Singh, and A.~Doucet.
\newblock Random finite sets and sequential {M}onte {C}arlo methods in
  multi-target tracking.
\newblock In \emph{Radar Conference, 2003. Proceedings of the International},
  pages 486 -- 491, sept. 2003.
\newblock \doi{10.1109/RADAR.2003.1278790}.

\bibitem[Vo and Ma(2006)]{Vo_and_Ma_2006}
B.N. Vo and W.K. Ma.
\newblock The {G}aussian mixture probability hypothesis density filter.
\newblock \emph{Signal Processing, IEEE Transactions on}, 54\penalty0
  (11):\penalty0 4091--4104, 2006.

\bibitem[Wei and Tanner(1990)]{Wei_and_Tanner_1990}
Greg C.~G. Wei and Martin~A. Tanner.
\newblock A {M}onte {C}arlo implementation of the {E}{M} algorithm and the poor
  man's data augmentation algorithms.
\newblock \emph{Journal of the American Statistical Association}, 85\penalty0
  (411):\penalty0 699--704, 1990.
\newblock ISSN 01621459.
\newblock \doi{Wei\%20and\%20Tanner,\%201990}.
\newblock URL \url{http://dx.doi.org/Wei\%20and\%20Tanner,\%201990}.

\bibitem[Whiteley et~al.(2010)Whiteley, Singh, and
  Godsill]{Whiteley_et_al_2010}
N.~Whiteley, S.~Singh, and S.~Godsill.
\newblock Auxiliary particle implementation of probability hypothesis density
  filter.
\newblock \emph{Aerospace and Electronic Systems, IEEE Transactions on},
  46\penalty0 (3):\penalty0 1437--1454, 2010.

\bibitem[Y{\i}ld{\i}r{\i}m et~al.(2012)Y{\i}ld{\i}r{\i}m, Jiang, Singh, and
  Dean]{Yildirim_et_al_2012b}
S.~Y{\i}ld{\i}r{\i}m, L.~Jiang, S.~S. Singh, and T.~Dean.
\newblock A {M}onte {C}arlo expectation-maximisation algorithm for parameter
  estimation in multiple target tracking.
\newblock In \emph{15th International Conference on Information Fusion 2012, to
  appear}. Fusion 2012, 2012.

\bibitem[Yoon and Singh(2008)]{Yoon_and_Singh_2008}
J.W. Yoon and S.S. Singh.
\newblock A {B}ayesian approach to tracking in single molecule fluorescence
  microscopy.
\newblock Technical Report CUED/F-INFENG/TR-612, University of Cambridge,
  September 2008.

\end{thebibliography}

\end{document}